\documentclass[twocolumn,showpacs,preprintnumbers,amsmath,%
amssymb,nofootinbib,floatfix]{revtex4}
\usepackage{graphicx}
\usepackage{dcolumn}
\usepackage{bm}
\usepackage{amssymb}
\usepackage{calrsfs}
\usepackage[hypertex]{hyperref}

\def \ts{\langle\theta_{s}^2\rangle}
\def \es{\bar \epsilon}
\def \d{\partial}
\def \ds{\displaystyle}

\begin{document}


\title{Angular, spectral, and time distributions of highest energy protons and \\
associated secondary gamma-rays and neutrinos propagating through \\
extragalactic magnetic and radiation fields}

\author{F.A.~Aharonian}
\altaffiliation{Dublin Institute for Advanced Studies, 31 Fitzwilliam Place, Dublin 2, Ireland}
\altaffiliation{Max-Planck-Institut f\"ur Kernphysik,
Saupfercheckweg 1, D-6917 Heidelberg, Germany}
\email{Felix.Aharonian@mpi-hd.mpg.de}

\author{S.R.~Kelner}
\altaffiliation{Moscow Institute of Engineering Physics, Kashirskoe sh.
31, Moscow, 115409 Russia}
\altaffiliation{Max-Planck-Institut f\"ur Kernphysik,
Saupfercheckweg 1, D-6917 Heidelberg, Germany }
\email{Stanislav.Kelner@mpi-hd.mpg.de}

\author{A.Y.~Prosekin}
\affiliation{Max-Planck-Institut f\"ur Kernphysik,
Saupfercheckweg 1, D-6917 Heidelberg, Germany}
\email{Anton.Prosekin@mpi-hd.mpg.de}

\date{\today}

\begin{abstract}
The angular, spectral and temporal features of the highest
energy protons and accompanying them secondary neutrinos and
synchrotron gamma-rays propagating through the intergalactic magnetic and
radiation fields are studied using the
analytical solutions of the Boltzmann transport equation
obtained in the limit of the small-angle and continuous-energy-loss approximation.

\end{abstract}

\maketitle

\section{Introduction}
Because of deflections in the interstellar and intergalactic magnetic fields,
the information about the original directions of cosmic rays pointing to their
production sites is lost. On the other hand, the isotropic flux of cosmic rays
is contributed, most likely, by a large number of galactic and
extragalactic sources. These objects represent different source populations
characterized by essentially different physical parameters -- age, distance,
energy budget, etc., as well as by different particle acceleration scenarios.
This makes extremely difficult the identification of sources of cosmic rays
based on the chemical composition and energy spectra of particles - two measurables characterizing
the "soup"\, (isotropic flux of cosmic rays) cooked over cosmological timescales. Fortunately,
at extremely high energies, $E \sim 10^{20}$ eV, the impact of galactic and extragalactic
magnetic fields on the propagation of cosmic rays becomes less dramatic,
which might result in large and small scale anisotropies of cosmic ray fluxes.
Thus, depending on the strength and structure of the (highly unknown) intergalactic magnetic field (IGMF),
the highest energy domain of cosmic rays may offer us a new astronomical discipline -
"cosmic ray astronomy". The extension of studies to energies
$10^{20}$eV and beyond enhances the chances of localization of particle accelerators
for two reasons. With an increase of
particle energy, the probability that a proton would penetrate through
the intergalactic medium (IGM) without significant deflections
in chaotic magnetic fields increases. Note that for IGMF much weaker than
$10^{-9}$G, the deflection angle can be quite small also for lower
energy protons ($\theta \propto B/E$). However at energies significantly below
$10^{20}$ eV, the deflection in galactic magnetic fields
becomes the dominant factor leading to the lost of information
about the original directions of particles (see, e.g., Ref.~\cite{Stanev}).

In the context of prospects of realization of
"cosmic-ray astronomy", there is a second {\it independent}
factor which gives strong preference to energies $10^{20}$ eV.
Particles of such high energies can arrive only from relatively nearby
accelerators located within 100~Mpc (see, e.g.,~\cite{Cronin2004}).
This dramatically (by orders of magnitude) decreases the number of
relevant sources of $\geq 10^{20}$eV protons contributing to the observed cosmic ray flux,
and correspondingly reduces the level of the diffuse background,
i.e. the (quasi) isotropic flux as a superposition of contributions
by unresolved discrete sources. Formally, one cannot {\it a priori}
exclude the possibility that the $10^{20}$ eV cosmic rays are contributed by a large
number of weak sources which cannot be detected individually. Alternatively,
the entire cosmic ray flux at such high energies can be dominated by contributions from
a few sources, especially given the tough requirements to the $10^{20}$ eV
proton accelerators~\cite{FAetal2002}. This excludes, in particular,
objects like ordinary galaxies, unless the galaxies provide highest energy cosmic rays through
transient events related to compact objects like Gamma Ray Bursts
(GRB)~\cite{GRBs}.

The propagation of cosmic rays in IGMF has been discussed in a number of
recent works (see, e.g., Refs. \cite{Dolag,Sigl,Globus,Kotera}). In these
studies different magnetized environments have been assumed and different
methods dealing with particle transport have be applied. Consequently, their
conclusions are quite different, the principal reason being
the different assumptions and approaches in the modeling of the IGMF. The main
purpose of our work is the study of features related to the transport of
particles, therefore we assume, following Ref.~\cite{Globus}, purely turbulent
and homogeneous IGMF. This not only makes the calculations simple and more
transparent (as long as it concerns the pure transport effects), but also seems
to be a feasible realization for the large scale structure of IGMF.

Whether we may identify the accelerators of extragalactic cosmic rays using the highest
energy protons is a question which largely depends on the strength of the large scale IGMF.
Even for the most favorable conditions for realization of the "proton astronomy", the latter will be
relevant to the nearby Universe, the accessible sources being limited within a sphere of radius 100~Mpc.
A different approach for localization of acceleration cites of
$10^{20}$ eV protons can be provided by observations of
gamma-rays and neutrinos produced
at interactions of these energetic particles with the 2.7 K
CMBR photons and magnetic fields in the proximity of the source, namely
within a region of a size of order of 10~Mpc - sufficiently large for effective interaction of protons
with 2.7 K photons through photomeson process and, at the same time,
still small for a significant deflection of protons
from their original directions. All short-lived particles of these interactions, as well as
the products of their decays (gamma-rays, neutrinos and electrons) are
produced at small angles relative to the initial directions of parent protons. In an magnetized
environment with $B \geq 10^{-9}$~G the electrons with typical energy exceeding $10^{19}$ eV
are predominantly cooled via synchrotron radiation with production
of high energy gamma-rays. The electrons emit synchrotron photons very quickly,
before any significant change of their direction in the surrounding chaotic magnetic field. Thus the
synchrotron photons will move essentially in the initial direction of the parent protons. Since the
protons, after they escape their production site (accelerator), move radially, the observer
will see an apparent compact (quasi-point like) gamma-ray source \cite{FA2002,SG_FA2005},
even though gamma-rays are produced in an extended region with angular size of order of
$\sim d/r \sim 5 \,(r/100\, \text{Mpc})^{-1}$~deg.
Note that the same is true if protons escape the source anisotropically, but are moving within
a narrow angular cone towards the observer. Otherwise, the observer will miss the source.

The favorable range of IGMF for realization of this scenario is
$10^{-9} - 10^{-8}$~G. In a stronger magnetic field, deflections of protons are
significant even at the first several Mpc scales. Thus, because of the small interaction
depth of undeviated protons, the point like source becomes very weak.

On the other hand, for IGMF much weaker than $10^{-9}$G
electrons are cooled predominantly via inverse Compton scattering,
thus the efficiency of synchrotron radiation drops dramatically.
This scenario which involves a pair cascade in the 2.7 CMBR and Extragalactic Background Radiation (EBL),
also leads to high energy gamma-rays. However, unless the field is much weaker than
$10^{-12}$G, the cascade electrons of relatively low (TeV) energies are thermalized,
thus the cascade leads to the formation of giant halos \cite{halo} and in this way contribute
to the diffuse extragalactic gamma-ray background radiation (see, e.g.,
Ref.~\cite{diffuse_gamma})
rather than to the formation of a discrete gamma-ray source (for a discussion of different regimes
of formation of cascades initiated by interactions of highest energy protons
with 2.7 K ~ CMBR, and their detectability from the direction of the cosmic ray source
see Ref.~\cite{SG_FA2}). The detection of the cascade component as a point like
or a slightly
extended source of gamma-rays initiated by interactions of ultrahigh energy protons
(after they escape the accelerator) with
2.7~K CMBR is possible in the case of extremely small
IGMF, $B \leq 10^{-15} \ \rm G$ (see, e.g., Ref.~\cite{Semi_Ner}).

The energy spectrum and flux of synchrotron radiation of secondary electrons
from photomeson interactions of protons with 2.7 K CMBR
have been studied in Ref.~\cite{SG_FA2005}.
The calculations have been limited by the first 10 Mpc range of propagation of protons,
assuming that at this stage protons propagate radially without significant deviations, and the
secondary electrons move along the same direction before they emit synchrotron photons.
While this approximation gives a correct estimate of the flux, it does not specify the angle
within which the radiation is confined. This approach ignores also the
non-negligible tails of distribution of synchrotron radiation formed at the later stages of
propagation and interactions of protons.

In the case of quasi-continuous operation of an extragalactic accelerator of
protons over timescales exceeding the typical delay time due to the
deflection in the magnetic field, the energy and angular distributions of
protons, as well as accompanying photons and electrons, can be accurately
described by the steady-state solutions of the transport equations.
Generally, this is the case of a continuous proton accelerator of age $T \geq
10^6$~yr. In the case of shorter activity of the source (an "impulsive
accelerator") or solitary events like gamma-ray bursts, relatively simple
analytical solutions of the arrival time distributions of protons, gamma-rays
and neutrinos can be obtained within an approximation when the energy losses of
protons are ignored. We consider the cases of\, "continuous"\, and "impulsive"\,
proton accelerations in Sections \ref{continuous} and \ref{impulsive},
respectively.

\section{\label{continuous} Steady state distribution functions}

The realization of the small-angle multiple scattering considerably simplifies
the description of propagation of protons through a scattering medium.
In particular, in the small-angle approximation the term
$\bm v\frac{\d f}{\d \bm r}$ of the Boltzmann transport equation can be
 presented in a form allowing analytical derivation of the steady state solution.
Because of smallness of the single scattering angle one can
write the elastic collision integral in the Fokker-Planck approximation.
To expand the distribution function into series in terms of the
single scattering angle one should have a smooth function of this angle.
This condition is satisfied if one neglects unscattered part of the distribution
function that has very sharp angle dependency. Such an approximation
is justified in the case of multiple scattering. 

The approach provides solutions that can be applied to the various cases which,
independent of the details of the scattering medium, are characterized only by
the average scattering angle per unit length $\ts$. The scattering process
depends on the particle energy, i.e. $\ts$ is a function of energy. During the
propagation through the medium between two scattering centers, the energy of
particles is gradually decreased due to different dissipative processes.
If the change of energy in each action of interaction is considerably smaller
than the initial energy, one can use the continuous energy
loss approximation. It should be noted that in the approach described here
the processes responsible for the scattering and the energy loss of particle
are not required to be the same. The particle scattering could have elastic
character and do not cause energy losses. On the other hand, the effect of
deflection of particles from their original direction due to the processes
responsible for energy losses might be negligibly small. This is the case of the problem
considered below. One can safely ignore the change of the direction of
primary particles as well as the production angles ($\theta \sim 1/\gamma$)
of the secondary products (gamma-rays, electrons, neutrinos) due to all relevant
processes including photo-meson and pair production, inverse
Compton scattering, synchrotron radiation.

The aim of this section is to derive distribution functions for protons and
accompanying them
secondary particles propagating through the galactic and extragalactic magnetic fields
for a spherically symmetric point source of protons. However, it is technically more
convenient to consider first a source emitting protons in a given (fixed) direction. In
this case we have a preferential direction along the infinitely narrow beam
emitted by the source.
Let us choose z-axis along this direction. Because of the scattering,
particles deviate from the initial course. To define the deviation we introduce angles
$\theta_x$ and $\theta_y$ between the direction of propagation $\bm n$ and
the coordinate planes YOZ and XOZ, respectively. If $\bm n$ is close to z-axis, the
angles $\theta_x$ and $\theta_y$ are small and can be treated as components
of two-dimensional vector $\bm \theta$ in the XOY plane, where the absolute value
of $\bm\theta$ corresponds to the deflection angle between $\bm n$ and the
z-axis. Then we can write $\bm n\approx(\theta_x,\theta_y,1-\frac{\bm \theta^2}{2})=
({\bm\theta},1-\frac{\bm \theta^2}{2})$.

The retention of the second-order term $\bm
\theta^2/2$ in the expansion of $n_z$ allows us to take into account the effects relating to
the elongation of the path like delay time, but does not give any considerable contribution to steady-state solution. Therefore we divide the problem into two sub-problems.
In the first part of the paper we solve the steady-state equation that
takes into account the energy
losses but ignores the elongation of particle trajectories. The results of these calculations
are relevant to the "continuous"\, source of protons and describe the energy and angular
distributions of protons and accompanying neutrinos and synchrotron radiation of
secondary electrons produced during the propagation of protons. In the second part of
the paper we calculate the distributions of arrival times of protons, neutrinos and gamma-rays
in the case of an "impulsive"\, source. In this case the arrival time delays directly depend
on the elongation of trajectory. The time-dependent solutions for distribution
functions presented in Section \ref{impulsive} are limited by the approximation in which
the energy losses of protons are neglected.

Thus, to derive the steady-state solution of the transport equation
we assume $\bm n\approx(\theta_x,\theta_y,1)=({\bm\theta},1)$ .
Let us denote by $\bm\rho=(x,y)$ the perpendicular
displacement in the plane XOY. For
a point source characterized by a monoenergetic and infinitely narrow
beam of protons emitted along the z-axis we obtain the
equation for Green function $G(\bm r,\bm \theta,E)$ of the Boltzmann steady-state transport equation
in the approximations of a small-angle multiple
scattering and continuous energy losses:
\begin{eqnarray}
\label{eq1}
\left(\frac{\d}{\d z}+{\bm\theta}\frac{\d}{\d {\bm\rho}}-
\frac{\ts}{4}\frac{\d^2}{\d \bm{\theta}^2}
-\frac{\d}{\d E}\es \right)G(\bm r,\bm \theta,E,E_{0})\nonumber \\
=\frac{1}{c}\delta(z)\delta(\bm \rho)\delta(\bm\theta)\delta(E-E_0).
\end{eqnarray}
Here we take into account that the particles are ultrarelativistic $\left\vert
\bm v \right\vert=c$. The solution of Eq.~(\ref {eq1}) is obtained in
Ref.~\cite{Eyges} for the propagation of charged particles passing through a
layer of matter. The features of this solution are comprehensively discussed in
Ref.~\cite{Remizovich}. Using the notations introduced in
Ref.~\cite{Remizovich}, the Green function can be written in the form:
\begin{eqnarray}
\label{eq2}
G(\bm r,\bm \theta,E,E_0)=
\frac{\delta(S(E,E_0)-z)}{c\es(E) \pi^2 \Delta} \nonumber \\
\times\exp\!\left(-\frac{A_1\bm \rho\,^2-2A_2\bm \theta \bm \rho
+A_3\bm \theta\,^2}{\Delta} \right),
\end{eqnarray}
where $S$ is the traveled distance that is uniquely related to the energy loss rate
${\es(E)=|dE/dz}|$:
\begin{equation}\label{eqcr}
S(E,E_0)=\int\limits_{E}^{E_0}\frac{dE'}{\es(E')} \ ,
\end{equation}
and
\begin{equation}\label{eqDel}
\Delta=A_1A_3-A_2^2.
\end{equation}
The $\delta$-function in Eq.~(\ref{eq2}) points to the fact that we neglect
the elongation of trajectory so the traveled distance is equal to $z$ as
if particles propagate strictly along z-axis. Taking the relation between energy
and $z$ into account, $A_i$ can be written in the following form:
\begin{equation}\label{eq3}
A_i(E_{0},z)=\int \limits_0^z \ts(z')(z-z')^{i-1}dz'.
\end{equation}

It is easy to recognize the physical meanings of the coefficients $A_1,A_2$ and
$A_3$; $A_1$ is the mean square deflection angle, $A_3$ is the mean square displacement,
and $A_2$ is the mean value of $\bm \theta \bm\rho$ at the distance $z$:
\begin{equation}
A_1=\langle\bm \theta^2\rangle_z,\quad A_2=\langle \bm \theta \bm
\rho\, \rangle_z,\quad A_3=\langle\bm \rho^2\rangle_z.
\end{equation}

For the treatment of the case of spherically symmetric point source of protons,
let us rewrite Green function in the form which is independent of choice of
the coordinate system. After the replacements
\begin{equation}
\bm \theta\rightarrow\bm n-\bm n_0,\quad \bm \rho\rightarrow\bm r-r\bm
n_0,\quad z\rightarrow r,
\end{equation}
where $\bm n_0$ is the direction of
the emission,
$\bm n$ is the direction of particle motion at the point $\bm
r$, we find
\begin{widetext}
\begin{eqnarray}\label{greenfunc}
G(\bm r,\bm n,\bm n_0,E,E_0)=
\frac{\delta(S(E,E_0)-r)}{c\es(E) \pi^2 \Delta}
\exp\!\left(-\frac{A_1(\bm r- r\bm n_0)\,^2-2A_2(\bm r- r\bm n_0)(\bm n
-\bm n_0)
+A_3\bm(\bm n-\bm n_0)^2}{\Delta} \right).
\end{eqnarray}
\end{widetext}
Performing integration over all directions of the vector $\bm n_0$ by the saddle
point method (see Appendix \ref{AGrFun}), we find
\begin{equation}\label{grs}
G_{sph}(r,\theta,E,E_0)=\frac{\delta(S(E,E_0)-r)}{c\es(E)r^2 \pi D}
\exp\!\left(-\frac{\,\,\theta^2}{D}\right),
\end{equation}
where
\begin{equation}
D=A_1-2\frac{A_2}{r}+\frac{A_3}{r^2}.
\end{equation}
Since we have spherically symmetric distribution, the Green function
depends only on the distance $r$ from the source and $\theta$ which is the angle
between the radius-vector from the source to the observation point and the
movement direction at this point.

We assume that the spherically symmetric source injects protons
into the intergalactic medium with a constant rate:
\begin{equation}
Q_p(\bm r,E)=cJ_{p}(E)\delta(\bm r).
\end{equation}
The substitution of this expression into
\begin{eqnarray}\label{eq4}
f(r,\theta,E)=\int Q(\bm r_0,E_0)\qquad\nonumber\\
\times G_{sph}(r-r_0,\theta,E,E_0)d\bm r_0 dE_0
\end{eqnarray}
gives
\begin{eqnarray}
f_p(r,\theta,E)=\frac{1}{\es(E)}
\int \limits_E^{\infty}\frac{ J_p(E_0)}{\pi r^2 D}
\exp\left(-\frac{\,\,\theta^2}{D}\right)\qquad\nonumber \\
\qquad\times\delta(S(E,E_0)-r)dE_0,
\end{eqnarray}
where $D$ can be written as
\begin{equation}\label{eq8}
D(E_0,r)=\frac{1}{r^2}\int \limits_0^r \ts(r') r'^2dr'.
\end{equation}

For the given energy and spatial distribution of protons we can calculate
number of secondary particles from the decays of $\pi$-mesons that
are produced at interactions between protons and 2.7 K CMBR photons.
To obtain the energy distributions of the secondary products - photons,
electrons and neutrinos, we use the approximation proposed in
Ref.~\cite{Kelner}. The energy of protons is ultrarelativistic so we can assume
that secondary particles initially move in the same direction as protons. The
distribution of second particles can be presented in the form
\begin{equation}\label{eq6}
Q(r,\theta, E)=\hat Q(f_p(r,\theta,E_p))
\end{equation}
where $\hat Q$ denotes an integral operator.
For example, for the energy distribution of protons $J_p(E)$, the energy
distribution of photons produced in photomeson interactions is
\begin{equation}
Q_{\gamma}(E_{\gamma})=\hat Q_{\gamma}(J_p(E_p)),
\end{equation}
where
\begin{equation}
\hat Q_{\gamma}(J_p(E_p))=
\int J_p(E_p)f_{ph}(\epsilon)w(E_{\gamma},E_p,\epsilon)dE_p d\epsilon \ .
\end{equation}
Here $f_{ph}$ is the distribution function of CMBR photons, $w$ is the
differential interaction rate of the $p \gamma$ interactions, namely, the
Bethe-Heitler pair production or photomeson production (see Ref.~\cite{Kelner}).
Since we are interested in the distribution of ultrarelativistic electrons that
weakly deviate in the magnetic field, we can apply the Green function given by
Eq.~(\ref{greenfunc})
to the source function given by Eq.~(\ref{eq6}). Note that $\hat Q$ acts
only on variable $E_p$ , therefore we can change the order of integration.
Tedious calculations (see Appendix \ref{ADisFun}) yield:
\begin{eqnarray}\label{eq7}
f_e(r,\theta,E_e)=\frac{1}{c\,\es_e(E_e)}
\int\limits_{E_e}^{\infty}dE_{e0} \hat Q_e \Bigg[\frac{1}{\es_p(E_p)}
\int \limits_{E_p}^{\infty}dE_{p0}\,\nonumber \\
\times\frac{J_p(E_{p0})}{r^2}
\frac{\exp\!\left(-\frac{\theta\,^2}{D_e+D_p}\right)}{\pi(D_e+D_p)}
\delta(S-r)\Bigg].
\end{eqnarray}
Here $S$ is sum of the distances traveled by proton to the point of interaction
with CMRB and traveled by electron from the point of production
to the point $r$:
\begin{equation}
S=S_p(E_p,E_{p0})+S_e(E_e,E_{e0}).
\end{equation}
The angular distribution of electrons in Eq.~(\ref{eq7}) is characterized by
\begin{equation}\label{eq9}
D_e=A_{e1}-2\frac{A_{e2}}{r}+\frac{A_{e3}}{r^2},
\end{equation}
where $ A_{ei}=A_{ei}(E_{e},E_{e0})$ have the same meaning as in
Eq.~(\ref{eq3}), and
\begin{equation}
D_p=\frac{1}{r^2}\int \limits_0^{r_0} \ts r'^2dr',
\end{equation}
where $r_0=S_p(E_p,E_{p0})$.

The main channel of production of gamma rays by HE electrons is synchrotron
radiation. Applying the modified Eq.~(\ref{sdis1}) for the spectrum of
synchrotron radiation in chaotic magnetic field to the distribution of electrons
given by Eq.~(\ref{eq7}), we find the angular and spatial distributions of gamma
ray. Let us describe the procedure as in Eq.~(\ref{eq6}) by
\begin{equation}
Q_s(r, \theta, E_{\gamma})=\hat Q_s(f_e(r,\theta,E_e)).
\end{equation}
In the general case the distribution function of gamma rays that are
characterized by the source function
$Q(\bm r,\bm n,E)$ and propagate through the absorbing
medium with extinction coefficient $k(E)$, is
\begin{equation}
f_{\gamma}(\bm r,\bm n,E)=\frac{1}{c}\int\limits_{0}^{\infty}Q(\bm r-\bm
n\tau,\bm n,E)e^{-k\tau}d\tau,
\end{equation}
where $\bm n$ is the direction of movement at the point $\bm r$
that coincides with the direction of emitting as the propagation of gamma
rays is rectilinear.
Since emitting electrons have ultrarelativistic energy we assume that the direction of radiation coincide with the electron direction, therefore the distribution function of synchrotron gamma rays for the
distribution of electrons given by Eq.~(\ref{eq7}) is
\begin{equation}
f_{\gamma}(r, \theta,E_\gamma)=\frac{1}{c}
\int \limits_0^{\infty}\hat Q_{s}
(f_e(\vert\bm r-\bm n\tau \vert,\theta_i,E_e))\,
e^{-k(E_\gamma)\tau}d\tau,
\end{equation}
where $\theta_i$ is the angle between $\bm n$ and $\bm r-\bm n \tau$, and
$\theta$ is the angle between $\bm n$ and $\bm r$.
It is convenient to perform the integration over $\tau$ using delta-function
in Eq.~(\ref{eq7}). For that we should change the order of integration so
the integration over $\tau$ becomes internal. Using the features of the
delta-function we find
\begin{equation}
\frac{\delta(\left\vert \bm r-\bm n\tau \right\vert-S)}
{\left\vert \bm r-\bm n\tau \right\vert^{2}}=\frac{1}{Sr}
\frac{1}{\sqrt{\left(\frac{S}{r}\right)^2-\sin^2\theta}}\sum\limits_{i=1}^{2}\delta(\tau-\tau_i) \ ,
\end{equation}
where $\tau_{1,2}=r(\cos\theta\mp \sqrt{\left(\frac{S}{r}\right)^2-\sin^2\theta})$.
Since the term corresponding to $\tau_2 $
does not contribute to large angles $\theta_i$ in exponent (see
Eq.~(\ref{eq7})) we keep only the term corresponding to $\tau_1$. After
performing relevant calculations we obtain:
\begin{eqnarray}
f_{\gamma}(\bm r,\theta,E_\gamma)=\hat Q_s \left\{\frac{1}{c^2\,\es_e(E_e)}
\int\limits_{E_e}^{\infty}dE_{e0}\hat Q_e\Bigg[\frac{1}{\es_p(E_p)}
\right.
\nonumber \\
\left.\times \int\limits_{E_p}^{\infty}dE_{p0}\tilde
f_p\eta\!\left(1-\frac{S}{r}\right)
\eta\!\left(\frac{S}{r}-\sin\theta\right)\Bigg]\right\},
\end{eqnarray}
where
\begin{equation}
\tilde f_p=
\frac{J_p(E_{p0})}{\frac{S}{r}\sqrt{\left(\frac{S}{r}\right)^2-\sin^2\theta}}\frac{
\exp\!\left(-\frac{\theta_1^2}{D_e+D'_p}\right)}{\pi r^2(D_e+D'_p)}
e^{-k(E_\gamma)\tau_1},
\end{equation}
$\eta$ is Heaviside function. The angle between $\bm n$ and $\bm
r-\bm n \tau$ is
\begin{equation}
\theta_{1}=\arccos\!\left(\frac{\sqrt{(S/r)^2-\sin^2\theta}}
{S/r} \right).
\end{equation}

\section{\label{PrSpec} The spectral and angular distributions of protons, photons and neutrinos}
\subsection{Protons}
Transport of protons substantially depends on the spatial distribution of
magnetic fields. The assumption of chaotically oriented magnetic cells is
usually used for estimates of the influence of IGMF
on cosmic-ray propagation (see, e.g., Ref.~\cite{Waxman}). The
spectral analysis of the correlation function of the magnetic field fluctuations
\cite{Tinyakov} provides a more appropriate and accurate treatment of the problem.
We use this approach for derivation of the mean square deflection angle.

UHE protons propagate large distances in IGMF without considerable deflections. Indeed, the evaluation of
the deflection angle $\delta\theta\simeq \lambda/r_g$ on the
correlation length $\lambda$ is
\begin{equation}
\delta\theta\simeq9\times10^{-3}\left( \frac{\lambda}{1 \text{\,Mpc}} \right)
\left( \frac{B}{10^{-9}\text{G}} \right)\left(
\frac{10^{20}\text{eV}}{E}\right)\text{rad},
\end{equation}
where $r_g=E/eB$ is the gyroradius of the ultrarelativistic particle. Therefore
change of direction of ultrahigh energy protons is small on the scale
$\lambda\simeq1$ Mpc. The proton energy can be assumed constant for this
scale. Then
 the proton motion in the magnetic
field is described by the equation
\begin{equation}\label{th1}
 \dot{\bm v}=\frac{ec}{E}\,[\bm v\times\bm B(\bm r)]\,.
\end{equation}
For ultrarelativistic particles $\bm v=c\bm n$, where
$\bm n$ is a unit vector. Rewriting the change in velocity over the time
$\Delta t$ in the form $\Delta \bm v=c\bm\theta$, we find
\begin{equation}\label{th2}
\bm \theta=\int\limits_t^{t+\Delta t}\frac{ec}{E}\,
[\bm n\times\bm B(\bm r(t))]\,dt\,.
\end{equation}
Since the deflection angle
is small, the trajectory of particle can be considered rectilinear in integration.

Now one should make an assumption about the statistical properties of the magnetic
fields. Here we assume that IGMF is a statistically isotropic and homogeneous. While
$\langle\bm\theta\rangle=0$ (since in this
case $\langle\bm B\rangle=0$), the mean square deflection is
\[
\langle\bm\theta^2\rangle=\left(\frac{ec}{E}\right)^2\int\!
\langle[\bm n\times\bm B_1][\bm n\times\bm B_2]\rangle\,dt_1\,dt_2\,
\]
\begin{equation}\label{th3}
=\left(\frac{e}{E}\right)^2\,(\delta_{\alpha\beta}-n_\alpha n_\alpha)
\int\!\langle B_{1\alpha}B_{2\beta}\rangle\,dz_1\,dz_2\,,
\end{equation}
where $\bm B_{1,2}=\bm B(\bm r(t_{1,2})) $. Here we switch to integration
over coordinates of particle, directing $z$-axis along $\bm n$.

Eq.~(\ref{th3}) includes the correlation function of the magnetic field
\begin{equation}\label{th4}
K_{\alpha\beta}(\bm r_1-\bm r_2)\equiv\langle B_\alpha(\bm r_1)B_\beta(\bm
r_2)\rangle\, .
\end{equation}
It depends only on the difference $(\bm r_1-\bm r_2)$ because we assume statistical
homogeneity of the magnetic field. The mean square of magnetic field is determined
as $\langle\bm B^2\rangle=K_{\alpha\alpha}(0)={\rm const}$.

To turn to the spectral description, $K_{\alpha\beta}$ should be written as a Fourier integral:
\begin{equation}\label{th5}
K_{\alpha\beta}(\bm r_1-\bm
r_2)=\int\!\tilde K_{\alpha\beta}(\bm k)
e^{i\bm k(\bm r_1-\bm r_2)}\,\frac{d^3k}{(2\pi)^3}\,.
\end{equation}
Since ${\rm div}\bm B=0$, then $K_{\alpha\beta}$ should satisfy the conditions
\begin{equation}
 \d K_{\alpha\beta}/\d x_{1\alpha}=0\,,\qquad
 \d K_{\alpha\beta}/\d x_{2\beta}=0\,.
\end{equation}
For function $\tilde K_{\alpha\beta}$ these conditions take on form
$\tilde K_{\alpha\beta}k_\alpha=0$, $\tilde K_{\alpha\beta}k_\beta=0$. Therefore,
if there is no preferential direction in space, $\tilde K_{\alpha\beta}$ has the
following structure:
\begin{equation}\label{th6}
\tilde K_{\alpha\beta}(\bm k)=
\frac12\bigg(\delta_{\alpha\beta}-\frac{k_\alpha k_\beta}
{\bm k^2}\bigg)\Phi(\bm k^2)\,\langle \bm B^2\rangle\,.
\end{equation}
Here constant factor $\langle \bm B^2\rangle$ is introduced such that
$\Phi(\bm k^2)$ meets the normalization condition:
\begin{equation}\label{th7}
\int\!\Phi(\bm k^2)\,\frac{d^3k}{(2\pi)^3}=
\frac{1}{2\pi^2}\int\limits_0^{\infty}\!\Phi(\bm k^2)\,k^2\,dk=1\,.
\end{equation}

It is convenient to change variables $z_1=z+\zeta/2$,
$z_2=z-\zeta/2$ in the integral in Eq.~(\ref{th3}). Assuming that the traveled
distance
$\Delta z$ is much greater than a characteristic scale on which the correlation
function tends to zero, one can extend the limits of integration over $\zeta$ to infinity. Meanwhile, the traveled distance should be smaller than the distance on which the proton loses
its energy appreciably. The integrand depends
only on $\zeta$, therefore the integration over $dz$ gives the length of integration
interval $\Delta z$. The mean square deflection angle is
proportional to the traveled distance, so the mean square deflection per unit
length is
\begin{equation}\label{th8}
 \langle\theta_s^2\rangle=\left(\frac eE\right)^2
 (\delta_{\alpha\beta}-n_\alpha n_\beta)
 \int\limits_{-\infty}^{\infty}\! K_{\alpha\beta}(0,0,\zeta)\,d\zeta.
\end{equation}

Integration of $K_{\alpha\beta}$ written in the form of Eq.~(\ref{th5}) over
$d\zeta$ gives $2\pi\delta(k_z)$ that allows us to calculate the integral over $dk_z$.
Then, the integral remains over the components of $\bf k$ perpendicular to $z$-axis.
Using Eq.~(\ref{th6}), we obtain
\begin{equation}\label{th9}
 \langle\theta_s^2\rangle=\frac12\left(\frac eE\right)^2\langle \bm B^2\rangle
 \int\frac{d^2k_\perp}{(2\pi)^2}\,\Phi(k_\perp^2)\,.
\end{equation}
The derived result can be written in the form
\begin{equation}\label{th10}
 \langle\theta_s^2\rangle=\frac{\pi}{2}\left(\frac eE\right)^2\langle \bm
B^2\rangle \Lambda\,,
\end{equation}
where
\begin{equation}\label{th11}
 \Lambda=\int\!\frac{d^3k}{(2\pi)^3}\,
 \frac1k\, \Phi(k^2)=
 \frac{1}{2\pi^2}\int\limits_0^{\infty}\!\Phi(\bm k^2)\,k\,dk\,.
\end{equation}
Taking into account Eq.~(\ref{th7}), the factor $\Lambda$ can be treated as the
mean value of $k^{-1}$, $\Lambda=\langle k^{-1}\rangle$. 

The calculation of
$\ts$ requires the spectral energy distribution of magnetic field. To obtain
final form
of $\ts$ we assume a power-law spectrum
\begin{equation}
\Phi(k^2)k^2 \sim\ \left\{ \begin{array}{ll}
\left(\frac{k_{0}}{k}\right)^{\alpha}, & k>k_0\\
\left(\frac{k}{k_0}\right)^{\beta}, & k<k_0
\end{array} \right.
\end{equation}
where $k_0$ is an absolute value of the wave vector corresponding to
the maximal scale of correlation $\lambda$: $k_0=2\pi/\lambda$. It gives
\begin{equation}
\langle\theta_s^2\rangle=\frac{(\alpha-1)(\beta+1)}{4\alpha\beta}\left(\frac eE\right)^2\langle \bm
B^2\rangle \lambda.
\end{equation}
Taking into account the turbulent character of IGMF that has $\langle\bm B\rangle=0$,
we take $\alpha=5/3$ which corresponds to the Kolmogorov turbulence.
The choice of the parameter $\beta$ is, to a certain extent, arbitrary. Here
we assume $\beta=1$ which leads to a simple expression for
the mean square deflection angle per unit length:
\begin{equation}\label{emsa}
\langle\theta_s^2\rangle=\frac{\lambda}{5}\left(\frac eE\right)^2\langle \bm
B^2\rangle \ .
\end{equation}
Because of uncertainties related to the
spectrum of IGMF, the numerical factor in Eq.~(\ref{emsa})
is somewhat different from the coefficients used in other papers
(see, e.g., Ref.~\cite{Waxman}).

Significant uncertainty in calculations of $\langle\theta_s^2\rangle$ is
related to the absolute value of the correlation length $\lambda$. It is expected to be
between 100~kpc and 1~Mpc, i.e. comparable to the
the characteristic distances between galaxies.
In the subsequent calculations we normalize the correlation
length to $\lambda=1$~Mpc, but the presented results can be easily
recalculated for any $\lambda$.

\begin{figure}
\begin{center}
\includegraphics[width=0.5\textwidth,angle=0]{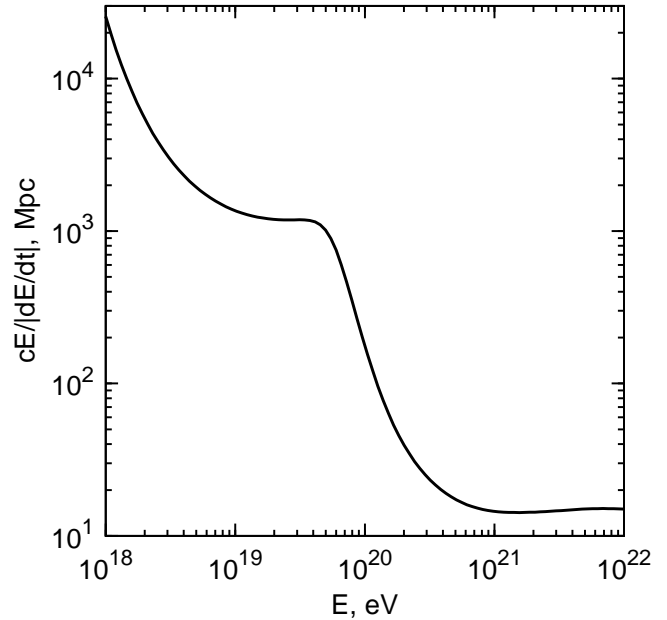}
\caption{\label{inv}The mean free path of protons in the intergalactic medium due to interactions with
photons of CMBR.}
\end{center}
\end{figure}

Since the propagation of protons in IGMF can be treated as a set of large number
of small chaotic deflections, the problem can be reduced to the diffusion in angle.
The diffusion coefficient $D(r,E_0)$ given by Eq.~(\ref{eq8})) contains
information
about the energy loss and influence of IGMF on propagation, and gives angular
distribution of protons at the given point.
Since there is a unique correspondence between the energy and $r$ (see
Eq.~(\ref{grs})) we can rewrite Eq.(\ref{eq8}) in terms of energy and energy
losses per unit length. Substituting Eq.(\ref{emsa}) into Eq.~(\ref{eq8}), we
obtain
\begin{equation}
D(E,E_0)=\frac{\eta}{r^2}\int\limits_{E}^{E_0}\frac{1}{E'^2}
\left(\int\limits_{E'}^{E_0}\frac{dE''}{\es(E'')}\right)^2\frac{dE'}{\es(E')},
\end{equation}
where
\begin{equation}
r=\int\limits_{E}^{E_0}\frac{dE'}{\es(E')},\qquad \eta=\frac{e^2\lambda}{5}
\langle \bm
B^2\rangle.
\end{equation}
The function $E/\es(E)$ based on results of Ref.~\cite{Kelner} and
implying the mean free path of protons in the intergalactic
medium due to the Bethe-Heitler pair-production and photomeson processes
at interactions with CMBR, is shown in Fig. \ref{inv}.

Note that for many scenarios described by Eq.~(\ref{eq1}) the same process is
responsible for both the angular scattering and the energy losses. But in the
case of propagation of protons in the intergalactic medium we deal with
two different processes: while the interactions with CMBR lead
to energy losses, the angular deflections are caused by multiple scattering on
magnetic inhomogeneities.

\begin{figure}
\begin{center}
\includegraphics[width=0.5\textwidth,angle=0]{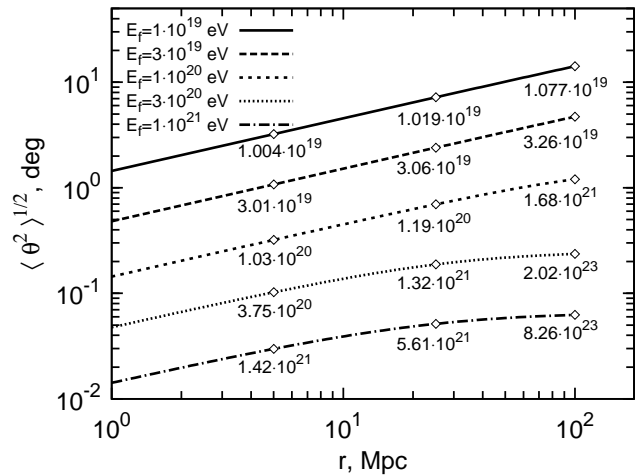}
\caption{\label{rms} The mean deflection angle of protons for the
fixed {\it observed} energy $E_f$ over the distance $r$. The numbers at the curves
indicate the energies which proton had at the distance $r$ from the observer.}
\end{center}
\end{figure}

The influence of energy losses on the angular distribution of protons can be
traced in Fig. \ref{rms}, where mean deflection angle of protons with observed
energies $E_f$ is shown as function of traveled distance $r$. As it is seen
from Fig. \ref{inv}, protons with energy smaller than $E=6\times10^{19}$ eV
do not suffer noticeable energy loses over the distances $\sim 100$ Mpc.
In this case the diffusion in angular space can be treated as a homogeneous random
walk that brings us to the dependence of the mean deflection angle on the
travel distance $\propto r^{1/2}$. For protons with initial energy higher than
the threshold of photomeson production, the energy of protons gradually decreases
which leads to deviation from this simple dependence. In particular, for the given
{\it observed} (final) energy $E_f$, this effect implies higher original energies, and
consequently smaller deflection angles at the initial parts of propagation. This
results in a weaker increase of the mean deflection angle with the
traveled distance in comparison with loss-free case. This effect is clearly seen
from analytical expressions, which is possible to obtain in the case of
constant energy loss rate $\left\vert \frac{dE}{dz}\right\vert \frac{1}{E}=b=\text{const}$:
\begin{equation}\label{ecl}
\langle\theta^2\rangle\ \sim \frac{r}{E^2_f}\left(\frac{\zeta^2-2\zeta+2(1-e^{-\zeta})}
{\zeta^3}\right)_{|\zeta=2br}.
\end{equation}
Expanding this expression into series in terms of powers of $r$ we obtain:
\begin{equation}\label{ecls}
\langle\theta^2\rangle\ \sim \frac{1}{E^2_f}\left( \frac{r}{3}-\frac{br^2}{6}+\cdots\right) \ .
\end{equation}
The first term does not depend on the value of $b$ and thereby describes the loss-free propagation.
The next term takes into account the energy losses and makes the dependence on the
distance $r$ weaker. While Eq.~(\ref{ecl}) approximately describes the behavior
of mean deflection
angle for the final energy $E_f \geq 10^{21}$ eV, the first term of
Eq.~(\ref{ecls}) describes the
case of $E_f \leq 6\times10^{19}$ eV.
In Fig.~\ref{rms} the mean deflection angle of protons is given for IGMF $B=1$ nG.
Since the dependence of the average deflection angle on the magnetic field is linear,
it is easy to produce plots for other magnetic fields.
\begin{figure*}
\begin{center}
\mbox{\includegraphics[width=0.5\textwidth,angle=0]{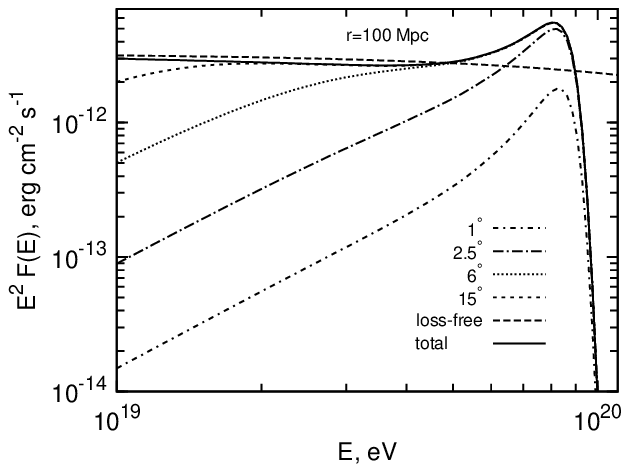}
\includegraphics[width=0.5\textwidth,angle=0]{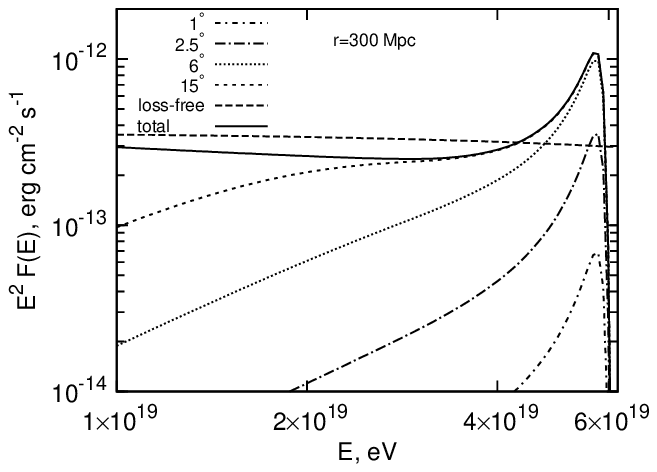}}
\caption{\label{fopr} Energy flux distribution of protons observed within different angles
for the source at the distance $r=100$ Mpc (left panel)
and $r=300$ Mpc (right panel). The initial spectrum of protons is assumed power-law
with an exponential cutoff at $E_0=3\times10^{20}$ eV, the IGMF is 1~nG.}
\end{center}
\end{figure*}

In order to indicate the evolution of the energy of protons during their propagation through the
2.7 K CMBR, in Fig.~\ref{rms} we indicate at the corresponding curves,
calculated for the fixed observed (final)
energies of protons, the energies which protons had at different distances from the observer.
For the fixed observed energies exceeding the threshold of photomeson
production, the calculated initial energies grow dramatically with the increase
of the distance, especially for $\geq 100$~Mpc. Therefore any deficit of protons
of such high energies in the initial spectrum
would results in the cutoff in observed spectrum at the corresponding energies.

The energy distributions of protons at distances $100$ Mpc and
$300$ Mpc are shown in Fig.~\ref{fopr} for the initial differential
energy spectrum $J_p(E)=J_{0} E^{-2}\exp(-E/E_0)$. The total
luminosity of the source in CRs with energy above $10^{9}$ eV is
taken $L=10^{44}$~erg/s. The upper dashed lines correspond
to the case when protons propagate in empty space; flux is determined
by the geometrical factor $1/r^2$. Solid line presents the case when
the deflections in the magnetic field are ignored. Comparison
of these two curves reveals two features: a bump and a sagging at lower energies.
Both features become more prominent with increasing of the distance.
The bump preceding the cutoff appears due to strong growth of energy losses
at the threshold of photomeson production (see Fig. \ref {inv}) that makes
particles to be accumulated in this energy region; the
sagging is a consequence of the energy losses due to the electron-positron
pair production (see, e.g., Ref.~\cite{Berezinsky}).

The approximation of continuous energy losses takes into consideration the
mean energy losses. In general it provides an acceptable accuracy but some features
connected with stochastic properties of interactions should be taken into account
for precise description of the spectrum in the cutoff region. The
fluctuations in the energy losses do have an impact on the form of the bump and the
cutoff in the observed spectrum of protons. It results, in particular, in a
smoother cutoff and a broader and lower-amplitude bump \cite{Cronin}
compared to the results calculated within the continuous energy losses approximation.

The impact of the magnetic field
leads to strong dependence of the energy distribution on the solid angle within which
the particles are detected. As it is seen in Fig.~\ref{fopr}, the flux of protons at highest energies
is concentrated along the direction to the source; the protons of lower energies
are scattered over large angles.

\subsection{Electrons}

The secondary gamma-rays and neutrinos are tracers of propagation of
protons in the intergalactic medium. The first generation gamma-rays
from photomeson processes are produced at extremely high energies
$E \geq 10^{19}$eV. They are effectively absorbed due to interactions with the
photons of CMBR and the Cosmic Radio Background (CRB) over distance $\sim 1 \
{\rm Mpc}$.
Because of the threshold effects, at energies below $10^{14}{\rm \, eV}$ the
efficiency of interactions with CMBR dramatically drops, but gamma-rays continue
to interact with the infrared and optical photons of the Extragalactic
Background Light (EBL). At these energies the mean three path of gamma-rays
increases sharply achieving, $\sim$100~Mpc at $E_\gamma \sim$10~TeV, and
$\sim$1~Gpc at $E \leq 200$~GeV (see, e.g., Ref.~\cite{EBL_abs}).
Thus, as long as we are interested in gamma-rays from the sources of
highest energy cosmic rays, the energy of gamma-rays should not significantly exceed 1~TeV.
In this energy band gamma-rays are produced through the electromagnetic cascade
initiated by the products of decays of short-lived mesons from the photomeson
interactions and, partly, by electrons from the Bethe-Heitler pair-production process.
For the development of an effective cascade the magnetic field should be
smaller than $10^{-10}$G. Even so, the observer can see the cascade gamma-rays
in the direction of the source only in the case of extremely small IGMF, $B \leq
10^{-15}$G. A collimated beam of gamma-rays of GeV--TeV energies is expected in
the case of magnetized intergalactic medium with $B \geq 10^{-9}$G. These
gamma-rays are produced through the synchrotron radiation of $E \geq 10^{19}$ eV
electrons.

Due to very large Lorentz factor of particles, we can assume
that secondary products from all interactions under consideration propagate strictly
in the direction of the parent particle. Therefore observed angular distribution
of gamma rays depends on the influence of IGMF on electrons that produce
these gamma rays. To observe the UHECR source in gamma rays it is necessary
that producing electrons are only slightly deflected in IGMF.

The almost rectilinear part of the path of electrons is much smaller than distances
traveled by protons and is comparable to the typical correlation length, $\lambda\simeq\
1$ Mpc. So the scattering of electrons takes place in almost homogeneous magnetic
field. But since direction of magnetic field have a random character the scattering
occurs in random directions. In case of electrons one can apply the formalism
of the multiple scattering to the random single scattering. Indeed, as have
been noted the distribution function should be smooth function of angle to
write the elastic collision integral in the Fokker-Planck approximation.
If all particles are scattered, as in the case of electrons, the distribution
function does not include a part with sharp angular distribution
corresponding to non-scattered particles.

\begin{figure}
\begin{center}
\includegraphics[width=0.5\textwidth,angle=0]{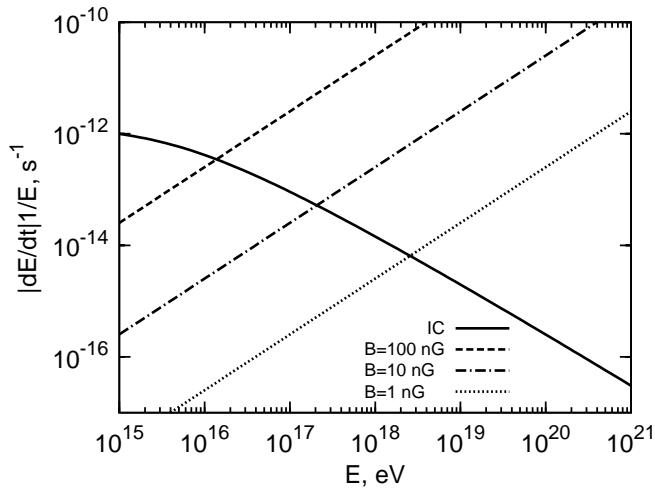}
\caption{\label{eloel}Energy loss rates of electrons due to
inverse Compton scattering on CMBR photons (solid line)
and synchrotron radiation in random magnetic field for $B=$1~nG, 10~nG, and 100~nG.
For electrons of energy $E \gtrsim 10^{19}$eV the inverse Compton scattering on
the radiowaves of
CRB becomes comparable or even can exceed the contribution of the Compton
scattering on CMBR, however for IGMF $B \gtrsim1$~nG the synchrotron radiation
remains the main
cooling channel.}
\end{center}
\end{figure}

To obtain the mean square deflection angle per unit length
we use expression for deflection angle of ultrarelativistic electron
traveled the path on which its energy has changed from the initial
energy $E'$ to the final energy $E$:
\begin{equation}\label{eq10}
\theta=\int\limits_{E}^{E'} \frac{1}{\es_er_g}dE'',
\end{equation}
where $\es_e$ is rate of energy losses due to synchrotron radiation and $r_g$
is gyroradius. Taking into account random field orientations we find
\begin{equation}
\ts=\frac{3}{2}
\frac{(mc^2)^4}{e^2E}\left( \frac{1}{E}-\frac{1}{E'}\right)
\left(\frac{1}{E}+\frac{2}{3}\frac{1}{E'}\right) \ .
\end{equation}
After substitution this equation into
Eq.~(\ref{eq3}) we find the coefficients
$A_{ei}$ of the diffusion coefficient $D_e$ in simple analytical forms:
\begin{equation}
A_1=\frac{5\alpha\beta}{180}\frac{1}{E^4}(7\xi^2+14\xi+9)(1-\xi)^2,
\end{equation}
\begin{equation}
A_2=\frac{\alpha\beta^2}{180}\frac{1}{E^5}(19\xi^2+22\xi+9)(1-\xi)^3,
\end{equation}
\begin{equation}
A_3=\frac{\alpha\beta^3}{180}\frac{1}{E^6}(12\xi^2+10\xi+3)(1-\xi)^4,
\end{equation}
where
\begin{equation}
\alpha=\frac{6}{5}\frac{(mc^2)^4}{e^2},\quad
\beta=\frac{9}{4}\frac{(mc^2)^4}{e^4B^2},\quad
\xi=\frac{E}{E'}.
\end{equation}
Eq.~(\ref{eq10}) can be written in the form
\begin{equation}\label{eqdefel}
\theta\approx0.008^\circ\frac{(1-\xi^2)}{B_{nG} E^2_{20}},
\end{equation}
where $B_{nG}$ is the magnetic field in units of nanoGauss (nG), $E_{20}$ is final energy in units
of $10^{20}$ eV, $\xi=E/E'$. Here the random orientations of the field are taken into
account. This expression allows us to estimate the threshold
of isotropization. Indeed, if the electron loses considerable part of its energy,
then $\xi\ll 1$ and deflection angle mostly depends on the final energy. The deflection
angle becomes quite large ($\sim 1$ radian) in the magnetic of
field $1$ nG when final energy is $E\approx 2\times 10^{18}$ eV. For greater
magnetic field the threshold of isotropization is shifted to the range of
lower energies. It should be noted that for magnetic fields $1-100$~nG
this threshold appears in the energy region where the energy losses due
to synchrotron radiation dominate over the inverse Compton scattering
(see Fig.~\ref{eloel}).
It means that inverse Compton scattering can be neglected for electrons under
consideration.

Let us estimate the energy of gamma rays produced by electrons with energy
exceeding the threshold of isotropization. Using modified Eq.~(\ref{sdis1}) for
energy distribution of synchrotron radiation in chaotic magnetic fields, we find
the energies of electrons that produce synchrotron gamma rays with energy
$E_{\gamma}$:
\begin{equation}\label{eq18}
E_{e}=1.23\times 10^{14}\sqrt{\frac{E_{\gamma}}{x B_{nG}}}.
\end{equation}
Here $E_e$ and $E_{\gamma}$ are given in units of eV,
$x$ is the dimensionless argument of distribution
function Eq.~(\ref{sdis2}). The latter has a maximum at $x\approx0.2291$
and exponentially decreases for large $x$ (see Eq.~(\ref{sdis3})). To make
sure that observed gamma rays are produced by electrons with energies greater
threshold of isotropization we should consider gamma rays with energies $E_{\gamma}\gtrsim 10^{9}$ eV. Indeed, electrons with energies corresponding to $x\gtrsim10$
in the Eq.~(\ref{eq18}) give exponentially small contribution into radiation
of gamma rays of the given energy. Therefore, assuming $x=10,$ we find that
the contribution of electrons with energies below threshold of isotropization $E_e\lesssim10^{18}$ eV into radiation of $E_\gamma=10^9$ eV gamma rays is
insignificant. According to Eq.~(\ref{eqdefel})
the product $BE^2$ is constant for the isotropization threshold. Since the same combination
enters in Eq.~(\ref{eq18}) the minimal energy of gamma rays produced
by the electrons under consideration does not depend on the magnetic field.

At interactions of protons with the intergalactic radiation fields
the ultrahigh energy electrons are produced via
two channels: pair production and photomeson production processes.
In the pair production process only a small $(\leq
2m_e/m_p)$ fraction of proton energy is converted to the secondary electrons.
For the magnetic field of order of nG or larger, the energies of these electrons appear below
the threshold of isotropization, thus they do not contribute to the gamma-ray emission
emitted towards the observer. The photomeson processes
lead to several non-stable secondary particles, such as $\pi$, $\eta$,
$K$ mesons, which decay into high energy gamma rays, neutrinos and electrons.
The electrons from the decays of these mesons are produced with energies \cite{Kelner}
exceeding the isotropization threshold.

In addition, a significant fraction of electrons is created at interactions of the first generation
("photomeson") gamma-rays with photons of CMBR and CRB. For the model of CRB
suggested by \cite{Protheroe}, the mean free path of gamma-rays of
$E \geq 10^{19}$eV is determined by the interactions with MHz radiowaves;
it is of order of several Mpc. Here we neglect by the interaction length assuming
that gamma rays interact with CRB immediately after their creation.
In this case the particle get additional deflection since it is treated
as electron all along. It results in broader angular distribution
of observed gamma ray in comparison with exact consideration.
The interaction of gamma rays with CRB photons of energy $\epsilon_R$
occurs in the regime $\epsilon_{R}E_{\gamma}/m_e^2 c^4 \gg1$. It means that the
most of the energy is converted to one of the two electrons.
The energy of gamma rays is higher than the energy
of electrons produced in the decays of mesons (see, e.g., Ref.~\cite{Kelner}).
Therefore electrons created by pair production process are more energetic than
electrons generated in the decays of nonstable products of photomeson processes.
Consequently, the pair-produced electrons result in higher flux of synchrotron
radiation than the direct ones from the meson decays.

\subsection{Gamma rays and neutrinos}
\begin{figure}
\begin{center}
\includegraphics[width=0.5\textwidth,angle=0]{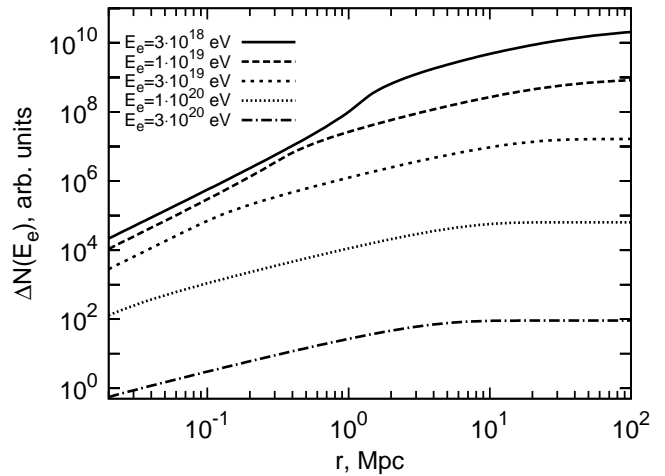}
\caption{\label{numoel}Number of electrons of energy $E_e$ located
inside a sphere of the radius $r$.}
\end{center}
\end{figure}

The apparent angular size of the synchrotron gamma-ray source
depends on the linear size of the emitter itself and the deflection
angles of the parent electrons. Both are defined by spatial and angular distributions
of electrons, respectively. In the case of spherically symmetric
source and small deflection angles of electrons $\theta_{def}$,
the source located at the distance $r$ with the gamma-ray emission region
of radius $d$, has an angular size
$\vartheta_{obs}\sim 2\frac{d}{r}\theta_{def}$. The case of isotropically
emitted gamma-ray source corresponds to $\theta_{def}\sim1$.
The linear size of the gamma-ray emitter can be evaluated from
Fig.~\ref{numoel}, where is shown the number of electrons of energy $E_e$ located
inside the sphere of radius $r$. The saturation that takes place at
large distances shows the absence of electrons in this region.
One can see from Fig.~\ref{numoel} that the size of the sphere, where the
electrons are located, decreases while the energy increases.
This is explained by the fact that protons producing electrons of such high
energies disappear due to energy losses. For energies below
$E_e=10^{19}{\rm\,eV}$ 
the electrons are not located in a definite region. The electrons
with energy below the thermalization threshold form an extended halo. These
electrons have energies at which the inverse Compton scattering losses 
dominates over the energy losses due to the synchrotron radiation. They initiate
electromagnetic cascades in the CMBR and EBL photon fields that eventually
results in a very extended GeV-TeV gamma ray emission.

\begin{figure*}
\begin{center}
\mbox{\includegraphics[width=0.5\textwidth,angle=0]{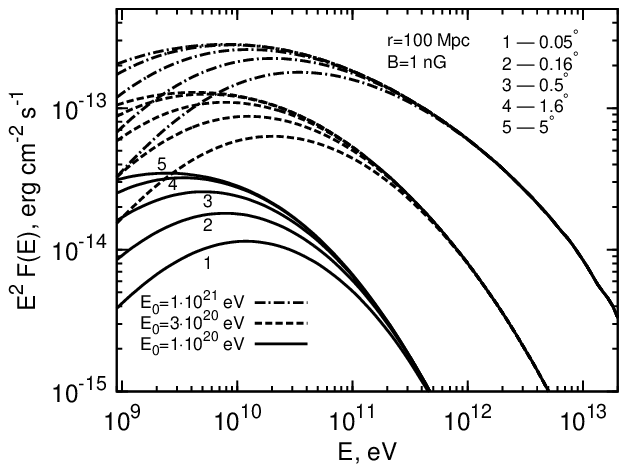}
\includegraphics[width=0.5\textwidth,angle=0]{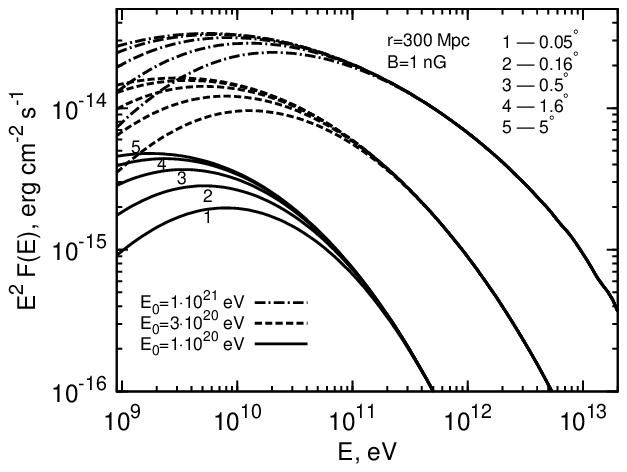}}
\caption{\label{fog}Flux distributions of gamma rays observed within different
angles in the direction of the source located at the distance $r=100$ Mpc (left panel)
and $r=300$ Mpc (right panel). Calculations correspond to the IGMF $B=$1~nG and
initial power-law distributions of protons with spectral index $\alpha=2$ and exponential
cutoffs at $E_0=10^{20}$eV; $3 \times10^{20}$eV, and $10^{21}$eV.
The total power of injection of protons into IGM is $10^{44} \ \rm erg/s$.}
\end{center}
\end{figure*}

\begin{figure}
\begin{center}
\includegraphics[width=0.5\textwidth,angle=0]{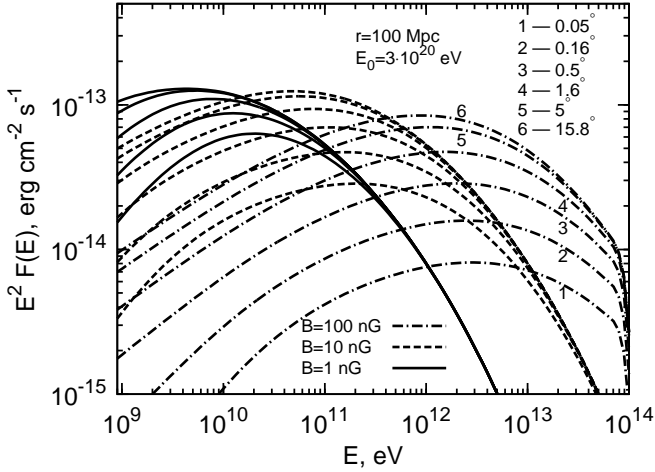}
\caption{\label{fogb} Flux distributions of gamma rays observed within
different angles in the direction of the cosmic ray source at the distance $r=100$ Mpc.
The calculations are performed for three different IGMF $B=$1~nG, 10~nG, 100~nG, assuming
an initial power-law distribution of protons with spectral index $\alpha=2$ an
exponential cutoff at $E_0=3\times 10^{20}$~eV. The total power of injection of
protons into IGM is $10^{44} \ \rm erg/s$.
The intergalactic absorption of gamma-rays due to interactions with EBL
is not taken into account.}
\end{center}
\end{figure}

\begin{figure*}
\begin{center}
\mbox{\includegraphics[width=0.5\textwidth,angle=0]{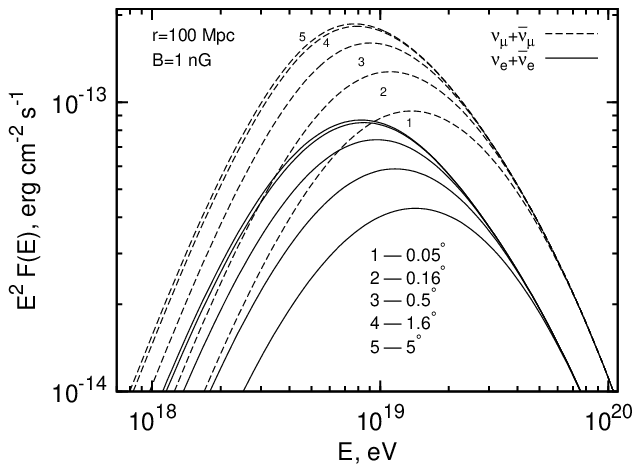}
\includegraphics[width=0.5\textwidth,angle=0]{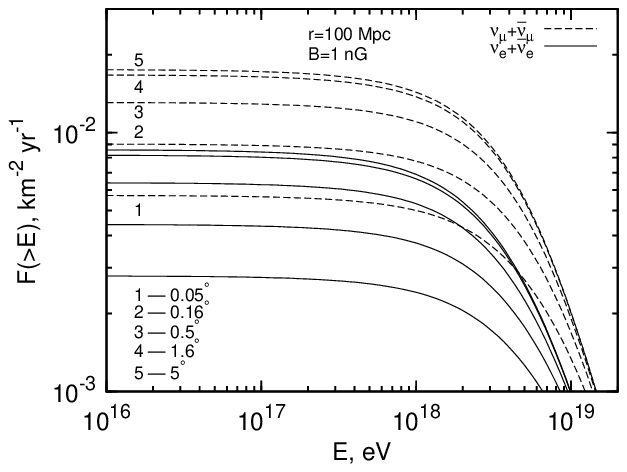}}
\caption{\label{fon} The fluxes of neutrinos observed within different angles
in the direction of the cosmic ray proton source located at the distance $r=100$ Mpc. The calculations are
performed for the initial power-law distribution of protons with spectral index
$\alpha=2$ and the exponential cutoff at $E_0=3\times10^{20}$ eV. The IGMF
$B=1$~nG, and the total power of injection of protons into IGM is $10^{44} \ \rm
erg/s$. Left panel is spectral energy distributions, right panel is integral
fluxes. }
\end{center}
\end{figure*}

\begin{figure}
\begin{center}
\includegraphics[width=0.5\textwidth,angle=0]{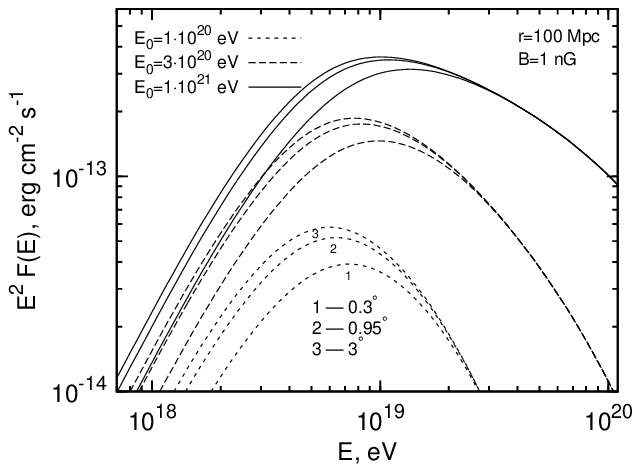}
\caption{\label{fneutcut} The spectral energy distributions of muon neutrinos
observed within different angles towards the source of cosmic ray protons
located at $r=100$ Mpc. The calculations are performed
for initial power-law distribution of protons with $\alpha=2$ and three
different values of the exponential cutoff: $E_0=10^{20}$eV,
$3\times10^{20}$eV, and $10^{21}$eV. }
\end{center}
\end{figure}

\begin{figure}
\begin{center}
\includegraphics[width=0.5\textwidth,angle=0]{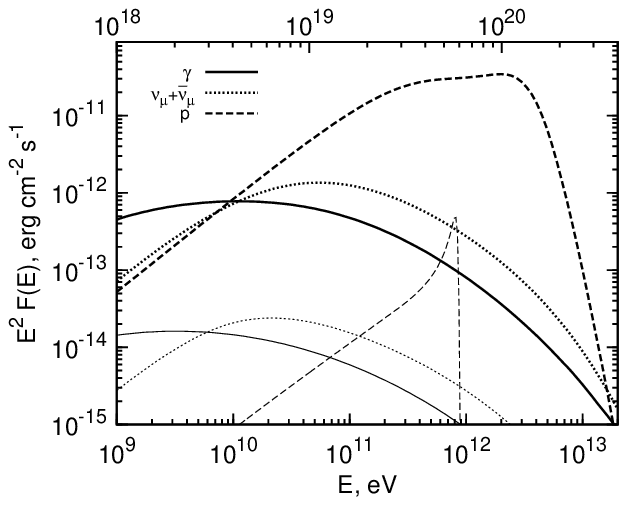}
\caption{\label{pres} The spectral energy distributions of gamma rays,
muon neutrinos and protons observed within the polar angle
$3^{\circ}$ from two identical source located at $r=30$ Mpc (thick lines) and
$r=300$ Mpc (thin lines).
The upper energy scale is for protons and neutrinos, the lower energy scale is
for gamma-rays.
The calculations are performed for the
initial power-law distribution of protons with spectral index $\alpha=2$, the
exponential cutoff $E_0=3\times 10^{20}$eV, and the
total power of injection into IGM $10^{44} \ \rm erg/s$. The IGMF is 1~nG.}
\end{center}
\end{figure}

\begin{figure}
\begin{center}
\includegraphics[width=0.5\textwidth,angle=0]{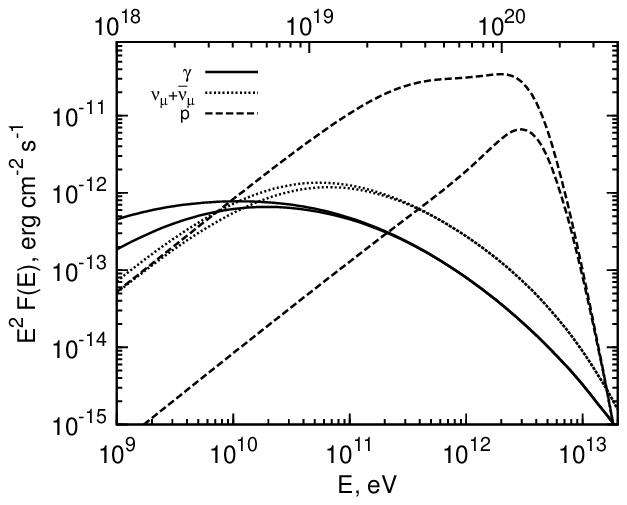}
\caption{\label{pres1} The spectral energy distribution of gamma rays, muon neutrinos and
protons observed within polar angles
$0.3^{\circ}$ and $3^{\circ}$ towards a source located at $r=30$~Mpc.
The parameters for the IGMF and the proton spectrum are the same
as in Fig. \ref{pres}.}
\end{center}
\end{figure}

The spectral energy distributions (SED) of gamma-rays, $E^2 F(E)$,
received within different angles are presented
in Fig.~\ref{fog}. The fluxes are calculated for the same initial proton energy distribution
used in Fig.~\ref{fopr}. Three series of curves for each of two distances (left and right panels)
correspond to different cutoff energies in the initial proton spectrum. One can
see that the cutoff energy has significant impact on the flux of
gamma rays; it increases the flux, shifts the maximum of SED towards higher energies,
and makes narrower the angular distributions. These features have a simple
explanation. The increase of the cutoff energy provides more secondary
electrons and extends the spectrum of electrons to more energetic region.
The latter leads to smaller deflections. It is interesting to note that although
the angular distribution of gamma rays is composed of deflections of both
protons and electrons, their angular distribution is more narrow compared to the
angular distribution of protons
(see, Fig.~\ref{fopr}). This is explained by the fact that the
main portion of gamma rays is produced in regions close to
the source by the highest energy protons which did not suffer
significant energy losses (see, Fig.~\ref{numoel}), while the multiple scattering
in IGMF contributes to the formation of the angular distribution of protons over
the entire path from the source to the observer.
Since the angular size of the gamma-ray source is determined by the
geometrical factor $d/r$, the distribution of gamma-rays from a source at
the distance $r=300$~Mpc is narrower than from an identical source located at
the distance $r=100$ Mpc. It is remarkable that at very high energies the
source becomes point-like. In particular, at energies above
$E_{\gamma}\approx5\times 10^{11}$~eV, the observer will see
the gamma-ray source located at the distance of 100~Mpc
within an angle smaller than $\theta_{obs}=0.1^{\circ}$.

Fig.~\ref{fogb} shows the impact of the IGMF strength on the flux distribution
of gamma rays. The increase of the magnetic field leads to the shift of
the maximum of SED to higher energies. In accordance with
Eq.~(\ref{eq18}), the shift of the synchrotron peak is
proportional to the strength of the magnetic field since the energy distribution
of electrons does not depend on the magnetic field. Finally note that
the increase of the magnetic field implies strong deflections which
leads to the reduction of the flux and widening of the
angular distribution of gamma-rays.

For the sources located beyond 100 Mpc, TeV gamma-rays
interact effectively with optical and infrared photons of
the Extragalactic Background Light (EBL).
The energy-dependent absorption of gamma-rays is characterized by the
optical depth $\tau_{\gamma \gamma}$ which
depends on the EBL flux and is proportional to the distance to the source.
Unfortunately the EBL flux contains quite large uncertainties,
especially at the mid and far IR wavelengths which are
most relevant to the gamma-ray energy band and the source distances discussed
in this paper. The impact of these uncertainties on the intergalactic
absorption of gamma-rays is discussed in Ref.~\cite{EBL_abs}. Even for the
minimum
EBL flux at infrared wavelengths, the absorption of TeV gamma-rays from sources beyond
100 Mpc can be significant; at multi-TeV energies the optical depth $\tau_{\gamma \gamma}$
exceed 1. Therefore the curves in Figs \ref{fog}, \ref{fogb} and \ref{fon}
should be corrected by multiplying the unabsorbed fluxes to
the factor $\exp(-\tau_{\gamma \gamma})$.

The decay of nonstable products of photomeson processes leads to the appearance of
extremely high energy electrons (positrons) and neutrinos (antineutrinos). Since the magnetic
field does not have an impact on neutrinos, the angular distribution of neutrinos is determined
only by the deflection of protons. This leads to more narrow angular distributions of neutrinos
compared not only to the distributions of protons (for the same reason described above
for gama-rays) but also compared to the distribution of gamma-rays
(because the gamma-ray distribution is additionally broadened due to deflections of electrons).
The left panel of Fig.~\ref{fon} shows SED of neutrinos and antineutrinos
received within different angles. The right panel of the figure presents
the integral fluxes of neutrinos. The impact of the cutoff energy
in the initial proton spectrum on the neutrino flux is demonstrated in Fig. \ref{fneutcut}.

For comparison, the spectral energy distributions of protons, gamma rays and
muon neutrinos are
shown together in Fig. \ref{pres} and Fig. \ref{pres1} for two distance to
the source - 30~Mpc and 300~Mpc.

\section{\label{impulsive} An impulsive source: arrival time distributions}

Let assume that at the moment $t=0$ an impulsive spherically-symmetric source
injects protons into the intergalactic medium. The multiple scattering of protons in the chaotic
magnetic field results in the deviation of the motion of particles
from the rectilinear propagation, therefore they arrive to the observer with significant time delays.
The arrival time of the proton moving with a speed $v_p$ over the path $S$ is
\begin{equation}\label{t1}
t=\frac{S}{v_p}=\frac{S}{c}+4.5\times10^{-4}\left(\frac{S_{\rm Mpc}}
{100}\right)\left(\frac{10^{18}}{E_{\rm eV}}\right)^2 \ {\rm s} .
\end{equation}
For ultrarelativistic protons the second term is negligible, therefore in calculations we adopt
$v_p=c$. In this paper we will study the distribution of the arrival-time delays
$\tau=t-r/c$ ignoring the energy losses of particles.

Let denote by $P(\tau,\zeta,r)\,d\zeta\,d\tau$
the probability that the proton with arrival direction in the interval
$(\zeta,\zeta+d\zeta)$ is detected at the distance $r$ from the source in
the time interval $(\tau,\tau+d\tau)$. Here $\zeta=\theta^2$, where
$\theta$ is the angle between the proton
direction at the point $\bm r$ and the
vector $\bm r$. It is assumed that $P$ satisfies to the
condition of normalization given by Eq.~(\ref{norm}).
The equation for the function $P$ for a pulse of radiation in the small-angle approximation
is obtained in Ref.~\cite{Alcock}. In Appendix \ref{ATimePr} we derive the
exact relation between $P$ and the standard distribution function
$f$, and obtain $P$ in a quite different (simpler) way than in
Ref.~\cite{Alcock}. Namely, our treatment of the problem is based on the
solution of equations written for the standard distribution function.

Following to Ref.~\cite{Alcock} we introduce the function $G$ which is
determined from the equation
\begin{equation}\label{t2}
P(\tau,\zeta,r)=\frac{c}{r^3\langle\theta^2_s\rangle^2}
\,G(x,y)\,,
\end{equation}
where the dimensionless parameters $x$ and $y$ are
\begin{equation}\label{t3}
 x=\frac{\zeta}{r\langle\theta^2_s\rangle}=\frac{\theta^2}
{r\langle\theta^2_s\rangle}\,,\qquad y=\frac{c\tau}
{r^2\langle\theta^2_s\rangle}.
\end{equation}
Function $G$ can be presented in the form of one-dimensional integral
\begin{equation}\label{t4}
 G(x,y)=\int\limits_{-\infty}^{\infty}\!\frac{ds}{2\pi}\,
\widetilde G(x,s)\,e^{isy}\,.
\end{equation}
Here
\begin{equation}\label{t5}
 \widetilde
G(x,s)=\frac{z}{j_1^{}(z)}\exp\Big(\!-x\frac{zj_0^{}(z)}{j_1^{}(z)}\Big)\,,
\end{equation}
where $z=\sqrt{s/(2i)}$, $j_0^{}$ and $j_1^{}$ are spherical Bessel functions:
\begin{equation}
 j_0^{}(z)=\frac{\sin z}{z}\,,\quad
j_1^{}(z)=\frac{\sin z}{z^2}-\frac{\cos z}{z}\,.
\end{equation}

The angular distribution of particles changes with time.
It can be shown, by using Eqs.~(\ref{t4}) and (\ref{t5}), that
\begin{equation}\label{t10}
 \langle \theta^2\rangle(\tau)=4c\tau/r\,,
\end{equation}
where
\begin{equation}\label{t10a}
\langle \theta^2\rangle(\tau)=\int\limits_0^{\infty}\! \theta^2G(x,y)\,dx\bigg/
\int\limits_0^{\infty}\! G(x,y)\,dx
\end{equation}
is the mean square deflection angle at the moment $\tau$.
Quite remarkably no model parameters enter in (\ref{t10}) in an explicit form.
Thus, the measurements of $\theta^2$ at different time periods allow
an estimate of the distance to the source. This is a nice feature,
because it could be the only channel of information about the distance to the source, if
the latter is not active anymore.

From Eqs. (\ref{t4}) and (\ref{t5}) follows that $G(x,y)=0$ at
$y<0$. We should note also the useful relation
\begin{equation}\label{t6}
 \widetilde G(x,s)=\int\limits_{-\infty}^{\infty}\!dy\,G(x,y)\,e^{-isy}\,,
\end{equation}
which allows us to obtain the moments of the function $G$:
\begin{equation}\label{t7}
m_n^{}\equiv \int\limits_{-\infty}^{\infty}\!dy\,y^n_{}\,G(x,y)=
i^n_{}\frac{\d^n}{\d s^n}\widetilde G(x,s)\Big|_{s=0}\,.
\end{equation}
Let's write down the first three moments:
\begin{eqnarray}
&\ds m_0^{}=3\,e^{-3x}\,,&\\
&\ds m_1^{}=\frac{3}{20}\,(1+2x)\,e^{-3x}\,,&\\
&\ds m_2^{}=\frac{3}{2800}\,(9+36x+28x^2)\,e^{-3x}.&
\end{eqnarray}
Correspondingly the mean values for $\langle y\rangle$ and
$\langle y^2\rangle$ are
\begin{eqnarray}
&\ds \langle y\rangle=\frac{m_1}{m_0}=\frac{1}{20}\,(1+2x)\,,\label{bary}&\\
&\ds \langle y^2\rangle
=\frac{m_2}{m_0}=\frac{1}{100}\,\left(\frac{9}{28}
+\frac{9}{7}\,x+x^2\right)\,.&
\end{eqnarray}
For the dispersion of distribution $\Delta$ and the ratio $\frac{\Delta}{\langle y\rangle^2}$
we have
\begin{equation}\label{t8}
\Delta\equiv \langle y^2\rangle-\langle y\rangle^2=\frac{1}{1400}\,(1+4x)\, ,
\end{equation}
and
\begin{equation}\label{t9}
 \frac{\Delta}{\langle y\rangle^2}=\frac27\,\frac{1+4x}{(1+2x)^2}\le \frac27\, .
\end{equation}
This implies that we deal with a rather narrow distribution. Rewriting
Eq.~(\ref{bary}) in the form
\begin{equation}
\frac{c\langle\tau\rangle}{r^2\langle\theta_s^2\rangle}=\frac{1}{20}
\left(1+\frac{2\theta^2}{r\langle\theta_s^2\rangle}\right)
\end{equation}
one can see that the measurement of $\langle\tau\rangle$
for the particles with different values of
$\theta$ allows to estimate $\langle\theta_s^2\rangle$.

Below we discuss two special cases of practical interest.

A. Detection of protons
with arbitrary arrival angles. This is the case discussed in Ref.~\cite{Alcock}.
In this case the distribution over $\tau$ is described as
\begin{equation}\label{t12}
 f_A^{}\equiv\int\limits_0^{\infty} \! P(\tau,\zeta,r)\,d\zeta=
\frac{4\pi^2c}{r^2\langle\theta_s^2\rangle}\sum\limits_{n=1}^{\infty}
(-1)^{n-1}n^2e^{-2\pi^2n^2y}\,.
\end{equation}
with mean values for $y$:
\begin{equation}\label{t12a}
\langle y\rangle=\frac{1}{12}\,,\quad \langle y^2\rangle=\frac{7}{720}\,,
\quad \Delta=\frac{1}{360}\,.
\end{equation}

B. Protons arriving along the radius-vector at the registration point. For this case,
substituting $x=0$ into Eq.~(\ref{t4}), we obtain
\begin{equation}\label{t13}
 f_B^{}\equiv P(\tau,\zeta=0,r)=-\frac{c}{r^3\langle\theta_s^2\rangle^2}
\sum\limits_{n=1}^{\infty}\frac{z_n^2}{j'_1(z_n)}\,e^{-2z_n^2y}\,,
\end{equation}
where $0<z_1<z_2<\cdots$ are the zeros of the function
$j_1^{}(z)$, located in the region $z>0$. 

\begin{figure}
\begin{center}
\includegraphics[width=0.3\textwidth,angle=-90]{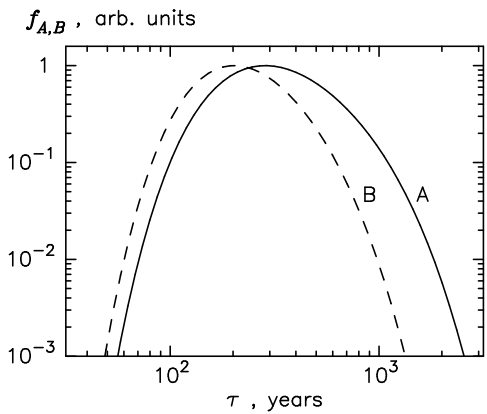}
\caption{\label{AB}\small The arrival time distributions for the cases
{\bf A} (solid line) and {\bf B} (dashed line) discussed in the text. The
distance to the source
is $r=10\ {\rm Mpc}$, the energy of protons $E=10^{20}\ {\rm eV}$, and the
strength of the magnetic field $B=1\ {\rm nG}$.
The curves are shown in arbitrary unites; for
convenience they are normalized to 1 at the points
of the maximum of distributions:
$\max(f_{A,B}^{})=1$.}
\end{center}
\end{figure}

The functions $f_A^{}$ and $f_B^{}$ corresponding to Eqs.~(\ref{t12}) and
(\ref{t13}) are shown in Fig.~\ref{AB}.

As it follows from Eq.~(\ref{t2}), the arrival time $\tau$ enters into $P$
only in the form of combination of the variable $y$. Since
\begin{equation}\label{t14}
 \lg y=\lg \tau-2\lg r-\lg\lambda-2\lg B+2\lg E +{\rm const}\,,
\end{equation}
the curves for other values of the relevant parameters, namely energy $E$,
magnetic field $B$, correlation length $\lambda$, and the distance to the source
$r$,
can be obtained by a simple shift along the $\tau$-axis.
However, it should be noted that Eq.~(\ref{t2}) is obtained in the approximation
of ignoring the energy losses of protons. Therefore for the large distances,
$r \geq 100$~Mpc, and especially for large energies, $E \geq 10^{20}$~eV,
Eq.~(\ref{t2}) overestimates the arrival time, given that the energy of protons
during their propagation significantly exceeds the energy at the registration point
(see Fig.~1). Therefore, for large distance Eq.~(\ref{t2}) should be
treated as an upper limit for the time delay. On the other hand, since
gamma-rays are produced at the very beginning of propagation of protons (within
10~Mpc or so), the curves calculated for a distance of order of 10~Mpc, provide
a quite accurate estimate for the arrival times of gamma-rays.

\section{Summary}

In this paper the angular, spectral and time distributions of UHE protons and
the associated secondary
gamma-rays and neutrinos propagating through the intergalactic radiation and magnetic fields
have been studied based on the relevant solutions of the
Boltzmann transport equation in the small-angle and continuous-energy-loss
approximations. A general formalism for the treatment of the steady
state distributions is provided in the form of relatively simple analytical
presentations. The treatment of the secondary products, in particular the
synchrotron gamma-radiation of electrons from photomeson interactions is reduced to the
consecutive application of the solutions which schematically can be presented as
\[Q_{p}\rightarrow f_{p}\rightarrow Q_{e}\rightarrow f_{e}
\rightarrow Q_{\gamma}\rightarrow f_\gamma.\]
Here $Q_i$ denotes a source function and $f_i$ denotes a distribution function.
$Q_p$ is specified as spherically symmetric source of protons. $Q_e$ is obtained
from distribution function of protons as the final product of photomeson
interactions using the results Ref.~\cite{Kelner}. Electrons generated in the
pair production
process of the first generation gamma-rays (from the decay of neutral
$\pi$-mesons)
are also included in $Q_e$. Finally, $Q_\gamma$ corresponds to the
synchrotron radiation of electrons with distribution function $f_e$ formed in
the chaotic magnetic field. We consider the case of strong magnetic field, $B
\geq 10^{-9}$~G,
when the electrons from photomeson interactions are cooled predominantly via synchrotron radiation.
Such strong magnetic fields prevent the development of pair cascades at highest energies, and, at the same time,
allow very effective conversion of the electromagnetic energy released at photomeson interactions
into synchrotron radiation. The latter peaks at GeV and TeV energies. The electromagnetic cascades are
developed at lower energies at which the suppression of the Compton cooling due to the
Klein-Nishina effect is becoming more relaxed. These sub-cascades are initiated basically by the electrons-positron pairs produced at the
inverse Bethe-Heitler process. However, because of deflections of low-energy electrons in chaotic IGMF,
the gamma-rays produced during the cascade development lose the directionality. Moreover, if the initial energy
distribution of protons extends to $10^{20}$eV, the electromagnetic energy released in photomeson interactions greatly exceeds the
energy supply from the Bethe-Heitler process. On the other hand, the synchrotron radiation produced by highest energy
secondary electrons not only provides an almost 100 \% effective conversion into gamma-rays, but also preserves
the initial direction of protons as long as the magnetic field does not exceed $10^{-7}$~G. Remarkably, while the main
fraction of synchrotron gamma-rays and the highest energy neutrinos is produced in the proximity of the source, namely within the first $\approx 10$~Mpc
of the initial path of protons, the latter continue to suffer deflections with an enhanced rate
(because of gradual decrease of energy during the propagation through the 2.7 CMBR), until they arrive to the observer.
Therefore the gamma-ray and neutrino distributions appear to be more narrow than the angular distribution of protons.

The distribution functions $f_p$ and $f_e$ are obtained by applying the Green
function of transport equation to the source functions $Q_p$ and $Q_e$, respectively.
The angular part of $f_p$ and $f_e$ is a normal (Gaussian-like) distribution,
the dispersion of which depends on the energy loss rate, the deflection angle
per unit length and the distance to the source. $f_\gamma$ is calculated by
integration along optical depth at different angles towards the source.

For specific realizations of the scenario of small-angle deflection of charged
particles, assuming that they move in a statistically isotropic and
homogeneous turbulent magnetic field with Kolmogorov spectrum, we considered the
IGMF in the interval from $10^{-9}$ to $10^{-7}$ Gauss and adopted 1~Mpc for the
correlation length. The propagation of protons is considered, as long as it
concerns the energy losses, as rectilinear with diffusion in angle. Transport
of electrons is considered in the homogeneous magnetic field with random
direction since their propagation length is of the order or less of $1$ Mpc.

Despite the small-angle scatterings, the related elongation of particle
trajectories causes significant delays of their arrival time.
The problem of propagation of particles can be described by
the steady state solutions if the lifetime of the source exceeds the delay times.
Otherwise the problem should be treated as a time-dependent propagation of particles
injected in the intergalactic medium by an "impulsive" source of extremely high energy
protons. This could be the case of solitary events like Gamma Ray Bursts or
short periods ($T \leq 10^5$~year) of enhanced activity of active galactic
nuclei. In this paper we discuss the case of an "impulsive" source,
ignoring the energy losses of protons. This approximation limits the
applicability of the derived time distribution functions to the relatively
nearby sources of protons located within 100~Mpc sphere of the nearby Universe.
On the other hand, since the bulk of synchrotron radiation of secondary
electrons is produced close the source, $R \lesssim 10\ {\rm Mpc}$, the
time-dependent solutions derived for protons, can describe quite accurately the
delayed arrival times of synchrotron photons from sources located at
cosmological distances.

The results presented in this paper for gamma-rays are valid for intergalactic magnetic
fields in a specific (but perhaps the most realistic) range between $B=10^{-9}$--$10^{-7}$G.
IMGF stronger than $10^{-7}$G would lead to large deflections of charged particles,
and thus violate the condition of small-angle approximation. On the other hand, IGMF weaker than
$10^{-10}$G would reduce dramatically the efficiency of the synchrotron radiation since in this case the
electrons are cooled predominantly via Compton scattering. The pair cascades
initiated by these electrons also lead to GeV and TeV gamma-ray emission, however
these cascades form giant (hardly detectable) halos around the sources, unless
the magnetic field is extremely weak, smaller than $10^{-15}$G.

The realization of the scenario of synchrotron radiation of secondary electrons
at the presence of a relatively modest magnetic field, $B \sim 10^{-9}$G or larger,
in the 10~Mpc proximity of the sources of highest energy cosmic rays,
has higher chances to be detected, given the compact (almost point like)
images at GeV and especially TeV energies, and the very high (10 per cent or more)
efficiency of conversion of the energy of protons to high energy synchrotron gamma-rays.
The fluxes of gamma-rays, protons and neutrinos shown in Figs.~\ref{fogb} --
\ref{pres1} are obtained assuming a power-law energy spectrum of protons with
$\alpha=2$ and total injection rate into IGM $\rm L_{\rm p} (\geq 1 GeV)=10^{44}
\ erg/s$. 
The expected gamma-ray fluxes are close to the sensitivities of {\it Fermi} LAT
at GeV energies and the sensitivity of
the Imaging Atmospheric Cherenkov telescope arrays at TeV energies. While the total power of production
of highest energy cosmic rays hardly can exceed, except for very powerful AGN, $10^{44} \rm erg/s$,
in the case of blazars with small beaming angles, the expected fluxes of gamma-rays could be
significantly higher. Indeed, in the case
of small deflections, the directions of injection of observed particles from
the source are close to the observational line. Therefore the results for
the spherically symmetric source remain valid also for the narrow jet
with solid angle $\Omega_{beam}=\pi\vartheta^2$, where
$\vartheta^2\gtrsim\langle\theta^2\rangle$. Then the required power of
the source (to be detected in gamma-rays) is reduced by a factor
$\Omega_{beam}/4\pi$ which may be significantly small.

\appendix
\section{\label{AGrFun} The Green function for spherically symmetric source}

The Green function for spherically symmetric point source is obtained by
integration of Eq.~(\ref{greenfunc}) over all directions of the vector $\bm
n_0$. Let's rewrite Eq.~(\ref{greenfunc}) in the following form:
\begin{equation}
G(\bm r,\bm n,\bm n_0,E,E_0)=\frac{\delta(S(E,E_0)-r)}{c\es(E) \pi^2 \Delta}
\exp\!\left(-\frac{A-\bm B \bm n_0}{\Delta}\right).
\end{equation}
Here
\begin{eqnarray}
A=2(A_1r^2-2A_2r+A_3)-2A_2r(1-\bm n_{r}\bm n),\nonumber\\
\bm B=2((A_1r^2-A_2r)\bm n_r+(A_3-A_2r)\bm n),
\end{eqnarray}
where $\bm n_r=\bm r/r$.
Since the directions of $\bm n$, $\bm n_0$ and $\bm n_r$ are close to each
other,
\begin{equation}
\bm B \bm n_0=\left\vert \bm B \right\vert\cos\theta_0
\approx\left\vert \bm B \right\vert \left(1-\frac{\theta_0^2}{2}\right).
\end{equation}
Performing integration by saddle point method we obtain:
\begin{equation}
\int \exp\!\left(-\frac{A-\bm B \bm n_0}{\Delta}\right)d\Omega_{\bm n_0}
\approx\frac{2\pi \Delta}{\left\vert \bm B
\right\vert}\exp\!\left(-\frac{A-\left\vert
\bm B \right\vert}{\Delta}\right).
\end{equation}
Taking into account
\begin{equation}
\bm n_r\bm n\approx 1-\frac{\theta^2}{2},
\end{equation}
the expressions for $A$ and $\left\vert \bm B \right\vert$ can be written:
\begin{eqnarray}
A=2Dr^{2}+\theta^2A_2r,\hspace{40pt} \nonumber\\
\left\vert \bm B \right\vert=2\sqrt{(Dr^{2})^{2}-\theta^2(A_1r^2-A_2r)(A_3-A_2r)}
\end{eqnarray}
where
\begin{equation}
D=A_1-2\frac{A_2}{r}+\frac{A_3}{r^2}.
\end{equation}
Expanding $\left\vert \bm B \right\vert$ into series in terms of $\theta$
to the second-order term in exponent and retaining the first term in denominator
we find
\begin{equation}
G_{sph}(r,\theta,E,E_0)=\frac{\delta(S(E,E_0)-r)}{c\es(E)r^2 \pi D}
\exp\!\left(-\frac{\,\,\theta^2}{D}\right).
\end{equation}
\section{\label{ADisFun}Distribution function of electrons}
\begin{widetext}
After changing the order of integration in
\begin{equation}
f_e(\bm r,\bm n,E_e)=\int \hat Q_e(f_p(\bm r_0,\bm n_0,E_p))G(\bm r-\bm
r_0,\bm n,\bm n_0,E_{e},E_{e0})d\bm r_0 \Omega_{\bm n_0}dE_{e0}
\end{equation}
we arrive at the following integral over directions of the emission of electrons
$\bm n_0$ at the point $\bm r_0$ and directions of $\bm r_0$:
\begin{eqnarray}
I=\int \exp\!\left(-\frac{(\bm n_0-\bm
n_s)^2}{D}\right)\hspace{170pt}\qquad\nonumber\\
\times\exp\!\left(-\frac{A_{1}(\bm r-\bm r_0-\left\vert \bm r-\bm r_0
\right\vert \bm n_0)^2-2A_{2}(\bm r-\bm r_0-\left\vert \bm r-\bm r_0 \right\vert
\bm
n_0)(\bm n-\bm n_0)+A_{3}(\bm n-\bm n_0)^2}{\Delta}\right)d\Omega_{\bm
n_s}d\Omega_{\bm n_0},
\end{eqnarray}
where
\begin{equation}
\bm r=r\bm n_r, \qquad \bm r_0=r_0\bm n_s,\qquad \bm r-\bm r_0=r'\bm n_0,
\end{equation}
$\Delta$ and $D$ are defined in Eq.~(\ref{eqDel}) and Eq.~(\ref{eq8}),
respectively.
Taking into account that all directions are close, the integral can be presented
in the following form:
\begin{equation}
I=\int e^{-(A-\bm B\bm n_s)}d\Omega_{\bm n_s}d\Omega_{\bm n_0},
\end{equation}
where
\begin{eqnarray}
A=X_0+X_1(1-\bm n_r \bm n_0)+X_2(1-\bm n_r\bm n)+X_3(1-\bm n_0\bm n),\nonumber\\
\bm B=Y_1\bm n_0+Y_2\bm n+Y_3 \bm n_r.\hspace{70pt}
\end{eqnarray}
Here we introduce the following notations:
\begin{eqnarray}
X_0=\frac{2}{D}+\frac{A_{1}}{\Delta}((r-r')^2+r_0^2), \quad X_1=\frac{2}{\Delta}(A_{1}r'-A_{2})r,
\quad X_2=\frac{2}{\Delta}A_{2}r,\quad X_3=\frac{2}{\Delta}(A_{3}-A_{2}r'),\nonumber\\
Y_1=\frac{2}{D}+\frac{2}{\Delta}(A_2-A_1r')r_0,\quad Y_2=-\frac{2}{\Delta}A_2r_0,
\quad Y_3=\frac{2}{\Delta}A_1rr_0.\hspace{70pt}
\end{eqnarray}
Since directions of $\bm n$, $\bm n_0$ and $\bm n_r$ are close, we can expand
$\vert \bm B \vert$ into series to the first order terms:
\begin{equation}
\vert \bm B\vert\approx\nu-\frac{Y_1Y_2}{\nu}(1-\bm n_0\bm n)-\frac{Y_1Y_3}{\nu}(1-\bm
n_0 \bm n_r)-\frac{Y_2Y_3}{\nu}(1-\bm n \bm n_r),
\end{equation}
where
\begin{equation}
\nu=Y_1+Y_2+Y_3.
\end{equation}
The integration of $I$ over $d\Omega_{\bm n_s}$ by saddle point method gives
\begin{equation}
I\approx\frac{2\pi}{\nu}\int e^{-(A-\vert B\vert)}d \Omega_{\bm n_0}
\end{equation}
To perform the integration over $d\Omega_{\bm n_0}$ by the same method, we
present the expression in the exponent in the following form:
\begin{eqnarray}
A-\vert \bm B\vert=\underbrace{X_0-\nu}_{Z_0}+\underbrace{\left(X_1+\frac{Y_1Y_3}{\nu}\right)}_{Z_1}(1-\bm
n_0 \bm n_r)+\underbrace{\left(X_2+\frac{Y_2Y_3}{\nu}\right)}_{Z_2}(1-\bm
n \bm n_r)+\underbrace{\left(X_3+\frac{Y_1Y_2}{\nu}\right)}_{Z_3}(1-\bm
n_0 \bm n)\nonumber\\
=\underbrace{Z_0+Z_1+Z_3+Z_2(1-\bm n \bm n_r)}_{\tilde A}-\underbrace{(Z_1\bm
n_r+Z_3 \bm n)}_{\tilde {\bm B}}\bm n_0=\tilde A-\tilde {\bm B}\bm n_0.\hspace{70pt}
\end{eqnarray}
Expanding $\vert \tilde{\bm B}\vert$ into series
\begin{equation}
\vert \tilde{\bm B}\vert\approx \mu-\frac{Z_1Z_3}{\mu}(1-\bm n\bm n_r),
\end{equation}
where
\begin{equation}
\mu=Z_1+Z_3
\end{equation}
we find
\begin{equation}
I\approx\frac{4\pi^2}{\nu\mu}e^{-(\tilde A-\tilde{\vert \bm B\vert})}.
\end{equation}
Making replacements of all notations by their actual values and taking into
account that
\begin{equation}
r'\approx r-r_0 \ ,
\end{equation}
we finally obtain
\begin{equation}
I=\frac{\pi^2\Delta D}{D_er^2+Dr^2_0}\exp\!\left(-\frac{\theta^2r^2}
{D_er^2+Dr^2_0}\right),
\end{equation}
where
\begin{equation}
D_e=A_{1}-2\frac{A_{2}}{r}+\frac{A_{3}}{r^2}
\end{equation}
and $\theta$ is the angle between $\bm n_r$ and $\bm n$. The Integration over
$r_0$ in the expression for $f_e(\bm r,\bm n,E_e)$ can be readily performed
because of $\delta$-function.
\end{widetext}

\section{\label{ATimePr} Distribution of arrival times in the case of "impulsive"\, source}

In the case of spherical symmetry the proton distribution function $f=f(t,r,\mu)$
depends on time $t$, distance to the source $r$, and the variable
$\mu=\cos\theta=(\bm n\bm r)/r$. Here $\bm n$ is a unit vector towards the
direction of the
proton speed. Let's normalize $f$ using the condition
\begin{equation}\label{ap1}
\int\limits_0^{\infty}\!dr\int\limits_{-1}^{1}\!d\mu\,r^2f(t,r,\mu)=1\, .
\end{equation}
Then $r^2f(t,r,\mu)\,dr\,d\mu$ is the probability that at the moment $t$ the proton
is located in the layer $(r,r+dr)$ and is moving in the direction within $(\mu,\mu+d\mu)$.
Let assume that propagation of a single particle is fixed, i.e.
the radius vector $\bm r_0(t)$ and the direction $\bm n_0(t)$ are certain functions of time.
For this particle, the distributions over $r$ and $\mu$ are described by $\delta$-functions:
\begin{equation}\label{ap2}
 f_0(t,r,\mu)=\frac1{r^2}\,\delta(r-r_0(t))\,\delta(\mu-\mu_0(t)),
\end{equation}
where $\mu_0(t)=(\bm n_0(t)\bm r_0(t))/r_0(t)$.

By averaging Eq.~(\ref{ap2}) over the ensemble of particles gives
the distribution function $f$:
\begin{equation}\label{ap3}
 f(t,r,\mu)=\langle f_0(t,r,\mu)\rangle\,.
\end{equation}

Let assume that for each particle $r_0(t)$ is a monotonically increasing function of time,
i.e. there are no particles in the ensemble with $\mu\le0$. Then the equation
$r=r_0(t)$ has a unique solution with $t=t_0(r)$, and thus one can write
\begin{eqnarray}\label{ap4}
 \delta(r-r_0(t))=\frac{1}{dr_0/dt}\,\delta(t-t_0(r))\nonumber\\
=\frac{1}{c\mu_0(t)}\,\delta(t-t_0(r)).
\end{eqnarray}
Since in Eq.~(\ref{ap2}) this expression is multiplied to
$\delta(\mu-\mu_0(t))$,
in the denominator one can replace $\mu_0(t)$ by $\mu$ and take the factor
$1/c\mu$ out of the integral.
This yields
\begin{eqnarray}\label{ap5}
\langle\delta(r-r_0(t))\,\delta(\mu-\mu_0(t))\rangle\nonumber\\
=\frac{1}{c\mu}\,\langle\delta(t-t_0(r))\,\delta(\mu-\tilde\mu_0(r))\rangle\,,
\end{eqnarray}
where $\tilde\mu_0(r)=\mu_0(t_0(r))$.

Function $\langle\delta(t-t_0(r))\,\delta(\mu-\tilde\mu_0(r))\rangle$
has the meaning of the probability distribution for $t$ and $\mu$.
Writing $t=\tau+r/c$, we obtain the probability distribution for
$\tau$ and $\mu$ at the point $r$:
\begin{equation}\label{ap6}
 P(\tau,\mu,r)=\langle\delta(\tau+r/c-t_0(r))\,
\delta(\mu-\tilde \mu_0(r))\rangle.
\end{equation}
From this equation follows that $P$ satisfies the condition of normalization
\begin{equation}\label{ap7}
\int\limits_0^{\infty}\!d\tau\int\limits_{-1}^{1}\!d\mu\,P(\tau,r,\mu)=1\,.
\end{equation}
Thus we arrive at the conclusion that the functions $P$ and $f$ are
related as
\begin{equation}\label{ap8}
 P(\tau,\mu,r)=c\mu r^2\,f(\tau+r/c,r,\mu)\equiv c\mu r^2\,f'(\tau,r,\mu)\,.
\end{equation}

The distribution function satisfies the equation
\begin{equation}\label{ap9}
 \frac1c\,\frac{\d f}{\d t}+(\bm n\nabla)f=I\,,
\end{equation}
where $I$ is the collision integral. In the case of spherical symmetry
\begin{equation}\label{ap10}
 (\bm n\nabla)f=\mu\frac{\d f}{\d r}+\frac{1-\mu^2}{r}\,\frac{\d f}{\d\mu}\,.
\end{equation}
Replacing the variables in Eq.~(\ref{ap9}) from $(t,r)$ to $(\tau=t-r/c,r)$
and presenting the collision integral in the Fokker-Planck approximation,
we obtain
\begin{eqnarray}\label{ap11}
 \frac{1-\mu}{c}\,\frac{\d f'}{\d\tau}+\mu\frac{\d f'}{\d r} +\frac{1-\mu^2}{r}
\,\frac{\d f'}{\d\mu}\nonumber\\
=\frac{\langle\theta^2_s\rangle}{4}\,\frac{\d}{\d\mu}
\Big((1-\mu^2)\frac{\d f'}{\d\mu}\Big).
\end{eqnarray}

In the case of an impulsive source and no scattering (i.e. $\langle\theta^2_s\rangle=0$)
the distribution function normalized according to Eq.~(\ref{ap1}) is
\begin{equation}\label{ap12}
 f(t,r,\mu)=2\pi\,\delta(\bm r-c\bm nt).
\end{equation}
It is convenient to rewrite Eq.~(\ref{ap12}) in the form
\begin{equation}\label{ap13}
 f(t,r,\mu)=\frac{1}{cr^2}\,\delta(t-r/c)\,\delta(\mu-1).
\end{equation}

In order to demonstrate that the generalized functions in the forms given
by Eqs. (\ref{ap12}) and (\ref{ap13}) are identical, one should multiply the right
parts of these equations to an arbitrary function $h(\bm r,\bm n)$ and integrate
over the space coordinates and the direction of the vector $\bm n$.
This implies that at $\langle\theta^2_s\rangle=0$
\begin{equation}\label{ap14}
 f'(\tau,r,\mu)=\frac{1}{cr^2}\,\delta(\tau)\,\delta(\mu-1).
\end{equation}
It is clear, from general physical considerations, that in the limit $r\to 0$
Eq.~(\ref{ap14}) is valid also at $\langle\theta^2_s\rangle\ne 0$. Therefore
Eq.~ (\ref{ap14}) can be treated as a boundary condition to Eq.~(\ref{ap11}) at
the point $r=0$.

An analytical solution of Eq.~(\ref{ap11}) is possible to derive in the
small-angle approximation. In the case of multiple scattering, the average angle
of deviation of at the distance $r$ is of order of
$(r\langle\theta^2_s\rangle)^{1/2}$. Therefore for $r\ll
1/\langle\theta^2_s\rangle$ one can use the small angle approximation.
Let $\mu=1-\zeta/2$, and let us denote function $f'(\tau,r,1-\zeta/2)$ by
$f'(\tau,r,\zeta)$. Assuming $\zeta\ll1$, from
Eq.~(\ref{ap11}) we obtain
\begin{equation}\label{ap15}
\frac{\d f'}{\d r}+\frac{\zeta}{2c} \,\frac{\d f'}{\d\tau}-
\frac{2\zeta}{r}\frac{\d f'}{\d\zeta}-\langle\theta^2_s\rangle
\frac{\d}{\d\zeta}\Big(\zeta\frac{\d f'}{\d\zeta}\Big)=0\,.
\end{equation}
To solve Eq.~(\ref{ap15}) we apply the Fourier transformation:
\begin{equation}\label{ap16}
\tilde f'(\omega,r,\zeta)=\int\limits_{-\infty}^{\infty}\!
f'(\tau,r,\zeta)\,e^{-i\omega\tau}\,d\tau\,.
\end{equation}
Function $\tilde f'$ satisfies the equation
\begin{equation}\label{ap17}
\frac{\d\tilde f'}{\d r}+\frac{i\omega\zeta}{2c} \,\tilde f'-
\frac{2\zeta}{r}\frac{\d\tilde f'}{\d\zeta}-\langle\theta^2_s\rangle
\frac{\d}{\d\zeta}\Big(\zeta\frac{\d\tilde f'}{\d\zeta}\Big)=0\,,
\end{equation}
and the boundary condition given by Eq.~(\ref{ap14}) becomes
\begin{equation}\label{ap18}
\tilde f'(\omega,r,\mu)=\frac{2}{cr^2}\,\delta(\zeta)\,,\quad r\to 0.
\end{equation}

Let's search the solution in the form
\begin{equation}\label{ap19}
\tilde f'=e^{-\zeta a(r)+b(r)}\,,
\end{equation}
where the functions $a(r)$, $b(r)$ do not depend on $\zeta$. Substituting
Eq.~(\ref{ap19})
in Eq.~(\ref{ap17}), we obtain the following ordinary differential equations:
\begin{eqnarray}
&\ds\frac{da}{dr}=\frac{i\omega}{2c}+\frac{2}{r}\,a-\langle\theta^2_s\rangle\,
a^2 \,,&\label{eqa}\\
&\ds\frac{db}{dr}=-\langle\theta^2_s\rangle\,a\,.&\label{eqb}
\end{eqnarray}
The solution to Eq.~(\ref{eqa}) is
\begin{equation}
a(r)=\frac{1}{r\langle\theta^2_s\rangle}\,\frac{zj_0^{}(z)}{j_1^{}(z)}\,,
\end{equation}
where $z=r\sqrt{\omega\langle\theta^2_s\rangle/(2ic)}$. The arbitrary constant
which appears in the solution is chosen requiring singularity
at the point $r=0$. At $r\to 0$ the function $a=3/r\langle\theta^2_s\rangle$.
The solution to Eq.~(\ref{eqb}) is
\begin{equation}
b(r)=\ln\!\left(\frac1{r^3}\,\frac{z}{j_1^{}(z)}\right)
+{\rm const}\,,
\end{equation}
therefore the function $\tilde f'$ is defined as
\begin{equation}
\tilde f'=C\,\frac1{r^3}\,\frac{z}{j_1^{}(z)}
\exp\!\left(-\frac{\zeta}{r\langle\theta^2_s\rangle}\,
\frac{zj_0^{}(z)}{j_1^{}(z)}\right).
\end{equation}
For determination of the constant $C$ one should use the boundary
condition given by Eq.~(\ref{ap18}). In the limit of small $r$,
and using the relation
\begin{equation}
\lim_{r\to 0}\left(\frac{\beta}{r}\,e^{-\zeta\beta/r}\right)=\delta(\zeta)\,,
\end{equation}
we find
\begin{equation}\label{eqa1}
\tilde
f'=C\,\frac3{r^3}\,\exp\!\left(-\frac{3\zeta}{r\langle\theta^2_s\rangle}\,
\right)=C\,\frac{\langle\theta^2_s\rangle}{r^2}\,\delta(\zeta)\, .
\end{equation}
Comparing Eqs. (\ref{eqa1}) and (\ref{ap18}), we obtain
$C=2/c\langle\theta^2_s\rangle$ and then using Eq.~(\ref{ap8})
we find $P$. In the small-angle approximation we can replace the factor
$\mu$ in Eq.~(\ref{ap8}) by unity. In order to compare our results with
the solution obtained in Ref.~\cite{Alcock}, we adopt
$C=1/c\langle\theta^2_s\rangle$, which is equivalent to
the change of the condition of normalization, namely instead of
Eq.~(\ref{ap7}) we use
\begin{equation}\label{norm}
\int\limits_0^{\infty}\! d\tau \int\limits_0^{\infty}\! d\zeta
P(\tau,r,\zeta)=1\,,
\end{equation}
where, because of rapid convergence, the upper limit of integration over $d\zeta$
is set infinity. In order to present the result in the form given by Eqs.
(\ref{t2}) --
(\ref{t4}), one should introduce, instead of the variable $\omega$, a new variable of integration
$s=\omega r^2\langle\theta^2_s\rangle/c$.

\section{Emissivity function of synchrotron radiation in random magnetic
fields}

\begin{figure}
\begin{center}
\includegraphics[width=0.4\textwidth,angle=-90]{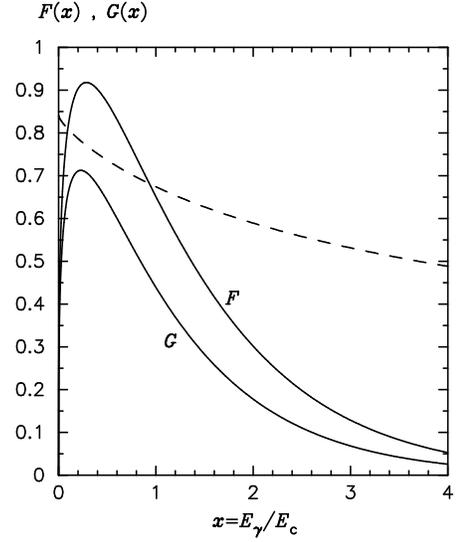}
\caption{\label{syndist}\small
The emissivity functions for synchrotron radiation $F(x)$ and $G(x)$.
The dashed line shows the ratio $G(x)/F(x)$.}
\end{center}
\end{figure}

For the case of chaotic magnetic fields one should average out the standard
formula for energy distribution of synchrotron radiation
\begin{equation}\label{sdis1}
\frac{dN_{\gamma}}{dE_{\gamma}dt}=\frac{\sqrt{3}}{2\pi}\frac{e^3B}{m_ec^2\hbar
E_{\gamma}}F\!\left(\frac{E_{\gamma}}{E_c}\right),
\end{equation}
where
\begin{equation}
F(x)=x\int\limits_x^{\infty}K_{5/3}(\tau)d\tau,\qquad E_c=\frac{3e\hbar
B\gamma^2}{2m_ec}\,,
\end{equation}
over directions of magnetic field. After taking the perpendicular
to velocity component of magnetic field $B_{\bot}=B\sin\theta$, where $\theta$
is angle between $\bm B$ and $\bm v$ we come to the following double integral:
\begin{equation}
G(x)=\int\sin\theta F\!\left(\frac{x}{\sin\theta}\right)\frac{d\Omega}{4\pi}=
\frac{1}{2}\int\limits_0^{\pi}F\!\left(\frac{x}{\sin\theta}\right)\sin^2\!\theta
d\theta.
\end{equation}
After changing the order of the integration it can be written as a single
integral
\begin{equation}
G(x)=x\int\limits_x^{\infty}K_{5/3}(\xi)\sqrt{1-\frac{x^2}{\xi^2}}d\xi,
\end{equation}
that can be expressed in terms of modified Bessel functions:
\begin{equation}\label{sdis2}
G(x)=\frac{x}{20}[(8+3x^2)(\kappa_{1/3})^2+x\kappa_{2/3}(2\kappa_{1/3}-3x\kappa_{2/3})],
\end{equation}
where $\kappa_{1/3}=K_{1/3}(x/2)$, $\kappa_{2/3}=K_{2/3}(x/2)$.
Note that while the function $F(x)$ has a maximum at $x=0.2858$
($\max F(x)=0.9180$), the maximum of the function $G(x)$ is shifted towards
smaller values: $x=0.2292$ ($\max G(x)=0.7126$). An alternative presentation
for $G(x)$ in terms of Whittaker's function has been derived in
Ref.~\cite{Crusius}. The functions $F(x)$ and $G(x)$,
as well as the ratio $G(x)/F(x)$ are shown in Fig.~\ref{syndist}.

Although the function $G(x)$ in Eq.~(D5) is presented in a quite compact and
elegant form, for practical purposes it is convenient to have
approximation which does not contain special functions. Here we propose
such approximations for $F(x)$ and $G(x)$ which provide an accuracy better
than 0.2~\% over the entire range of variable $x$:
\begin{eqnarray}\label{sdis4}
\tilde F(x)\approx2.15x^{1/3}(1+3.06x)^{1/6}\nonumber\\
\times\frac{1+0.884x^{2/3}+0.471x^{4/3}}
{1+1.64x^{2/3}+0.974x^{4/3}}e^{-x}\,,
\end{eqnarray}
\begin{equation}\label{sdis3}
\tilde
G(x)\approx\frac{1.808x^{1/3}}{\sqrt{1+3.4x^{2/3}}}\frac{1+2.21x^{2/3}+0.347x^{
4/3}}
{1+1.353x^{2/3}+0.217x^{4/3}}e^{-x}\,.
\end{equation}


\begin{thebibliography}{99}
%
\bibitem{Stanev} T. Stanev, Astrophys.J., {\bf 479}, 290 (1997)
%
\bibitem{Cronin2004} J.W. Cronin, Nuclear Physics B, {\bf 138}, p. 465 (2004).
%
\bibitem{FAetal2002} F.A. Aharonian, A.A. Belyanin, E.V. Derishev, V.V. Kocharovsky, Vl. V.; Kocharovsky, Vl. V.,
Physical Review D, {\bf 66}, id. 023005 (2002)
%
\bibitem{GRBs} M. Vietri, ApJ {\bf 453}, 883 (1995); E. Waxman,
Phys. Rev. Lett. {\bf75}, 386 (1995); M. Milgrom and V. Usov, ApJ {\bf 449}, L37 (1995)
%
\bibitem{Dolag}K. Dolag, D. Grasso, V. Springel, and I. Tkachev, J. Cosmol. Astropart. Phys. 01 (2005) 009
%
\bibitem{Sigl}G. Sigl, F. Miniati, and T. A. En\ss lin, Phys. Rev. D 70, 043007 (2004)
%
\bibitem{Globus}Globus, N., Allard, D., \& Parizot, E., A\&A, 479, 97 (2008)
%
\bibitem{Kotera}Kotera, K. \& Lemoine, M. 2008a, Phys. Rev. D, 77, 023005
%
\bibitem{FA2002} F.A Aharonian, MNRAS {332}, 215 (2002)
%
\bibitem{SG_FA2005}S. Gabici and F. A. Aharonian, Phys. Rev. Lett. {\bf95},
251102 (2005)
%
\bibitem{halo}F. A. Aharonian, P. S. Coppi, and H. J. V\"olk, ApJ {\bf 423}, L5 (1994)
%
\bibitem{diffuse_gamma} V. S. Berezinsky, A. Yu. Smirnov, Astrophys. Space Sci.
{\bf 32} 461 (1975); P.S. Coppi, F.A. Aharonian,
ApJ {\bf 487}, L9 (1997); O. Kalashev, D.V. Semikoz, G. Sigl, Phys.
Review D {\bf 79}, 063005 (2009)
%
\bibitem{SG_FA2} S. Gabici, F.A. Aharonian, Astrophys. Space Sci. {\bf 309} 465 (2007)
%
\bibitem{Semi_Ner} A. Yelyiv, A.Neronov, D.V. Semikoz, Phys. ReV {\bf D80}, 023010 (2009)
%
\bibitem{Eyges}L. Eyges , Phys. Rev. {\bf 74}, 1534 (1948).
%
\bibitem{Remizovich}V. S. Remizovich, D. B. Rogozkin, and M. I. Ryazanov, Charged Particles Path-Length Fluctuation (Energoatomizdat, Moscow, 1988).
%
\bibitem{Kelner} S.R. Kelner, F.A. Aharonian, Phys. Rev. D {\bf78}, 034013 (2008).
%
\bibitem{Waxman} E. Waxman and J. Miralda-Escude, Ap. J. 472, L89 (1996).
%
\bibitem{Tinyakov} P. G. Tinyakov and I. I. Tkachev, Astropart. Phys. {\bf24},
32 (2005).
%
\bibitem{Berezinsky}V. S. Berezinsky and S. I. Grigor'eva, Astron. Astrophys.
{\bf 199}, 1 (1988)
%
\bibitem{EBL_abs} F.A. Aharonian, A.N. Timokhin, A.V. Plyasheshnikov, Astron. Astrophys. {\bf 384},
384 (2002)
%
\bibitem{Protheroe}R. J. Protheroe and P. L. Biermann, Astropart. Phys. {\bf6}, 45 (1996)
%
\bibitem{Cronin}F.A. Aharonian and J.W. Cronin, Phys. Rev. D {\bf50}, 1892 (1994)
%
\bibitem{Alcock}Alcock, C. and Hatchett, H., ApJ, 222, 456 (1978)
%
\bibitem{Crusius}Crusius, A., and Schlickeiser, R., A\&A, 164, L16 (1986)
\end{thebibliography}
\end{document}